\documentclass{elsart}
\usepackage{graphicx,natbib,amssymb}
\usepackage{setspace}
\makeatletter
\def\elsartstyle{%
	\def\normalsize{\@setfontsize\normalsize\@xiipt{14.5}}
	\def\small{\@setfontsize\small\@xipt{13.6}}
	\let\footnotesize=\small
	\def\large{\@setfontsize\large\@xivpt{18}}
	\def\Large{\@setfontsize\Large\@xviipt{22}}
	\skip\@mpfootins = 18\p@ \@plus 2\p@
	\normalsize
}
\makeatother

\def\url#1{{\ttfamily\def\/{/\discretionary{}{}{}}#1}}

\begin{document}

\begin{frontmatter}
\title{Spectroscopic followup of three bright halo stars selected from SDSS and GALEX
photometry}

\begin{center}
\large
T. \c{S}ahin$^{\mathrm{1}*}$,
D. L. Lambert$^{\mathrm{2}}$,
C. A. Prieto$^{\mathrm{2,3,4}}$\\[12pt]

\small\itshape
$^1$Akdeniz University, Faculty of Science, Space Science and Technologies Department, 07058, Antalya, Turkey\\
$^2$Department of Astronomy and The W.J.  McDonald Observatory, University of
Texas, Austin, TX 78712, USA\\
$^3$Instituto de Astrof\'{\i}sica de Canarias, 38205, La Laguna, Tenerife, Spain
$^4$Departamento de Astrof\'{\i}sica, Universidad de La Laguna, 38206, La Laguna, Tenerife, Spain

\end{center}

\thanks[email]{E-mail: timursahin@akdeniz.edu.tr}

\begin{abstract}

We aim to reveal the nature of the Sloan Digital Sky Survey (SDSS) stars: SDSS J100921.40+375233.9, SDSS J015717.04+135535.9, and SDSS J171422.43+283657.2, showing apparently high NUV excesses for their $g - z$ colors, as expected for extremely low-metallicity stars.
High resolution (R=60\,000) spectra of the stars with a wide wavelength coverage were obtained to determine their chemical compositions with the Tull echelle spectrograph on the 2.7 m telescope at the McDonald Observatory. We derived the spectroscopic parameters $T_{eff}$\,=5820$\pm$125\,K, $\log g$\,=\,3.9$\pm$0.2, and $\xi_t$\,=\,1.1$\pm$0.5\,km s$^{-1}$ for SDSS J100921.40+375233.9, $T_{eff}$\,=6250$\pm$125\,K, $\log g$\,=\,3.7$\pm$0.2, and $\xi_t$\,=\,4.0$\pm$0.5\,km s$^{-1}$ for SDSS J015717.04+135535.9, and $T_{eff}$\,=6320$\pm$125\,K, $\log g$\,=\,4.1$\pm$0.3, and $\xi_t$\,=\,1.5$\pm$0.5\,km s$^{-1}$ for SDSS J171422.43+283657.2, and elemental abundances were computed for 21 elements for J100921 and J171422 and for 19 elements for J015717 for the first time. We find metallicities of [Fe/H]$=-1.30$, $-0.94$, and $-0.80$ for SDSS J100921.40+375233.9, J015717.04+135535.9, and J171422.43+283657.2, respectively. On the basis of calculated abundance ratios for J171422.43+283657.2 and J015717.04+135535.9, we also report that these two program stars have the expected composition of main-sequence halo turnoff stars, but with low-$\alpha$ abundances, i.e., the [$\alpha$/Fe] ratio is $\approx$0.0 for J171422.43+283657.2 and $\approx$0.1 for J015717.04+135535.9. The latter one shows typical halo or thick-disk $\alpha$-element abundances, but has a substantial rotational line broadening and $v$ sin$i$=40 $\pm$ 0.5 km s$^{-1}$

\end{abstract}

\end{frontmatter}

\section{Introduction}
\label{intro}

Stars with extremely low iron abundances are rare, and only a handful are  known at
[Fe/H]$<-3.5$ \footnote{Standard notation is used for quantities [X] where
[X]=$\log$(X)$_{\rm star}$-$\log$(X)$_\odot$.}  (Beers \& Christlieb 2005; Sch{\"o}rck
et al. 2009). Therefore, any efforts to identify additional candidates in this regime are
important. Metal-poor stars exhibit a distinct UV excess compared to higher-metallicity
stars, as metal (mainly iron) line opacity becomes smaller. UV-excesses have long been used to identify halo stars in the
solar neighborhood (see, e.g., Eggen, Lynden-Bell \& Sandage 1962), and with the advent
of the Galaxy Evolution Explorer ({\sc GALEX}, Morrissey et al. 2007), UV fluxes are
now available for sources down to at least $m_{AB}<20$  over more than 25\,000 square
degrees of the sky, presenting us with a new opportunity to search for low-metallicity
objects.

 \begin{figure*}[!ht]
 \centering
 \includegraphics*[width=14cm,height=13cm,angle=0]{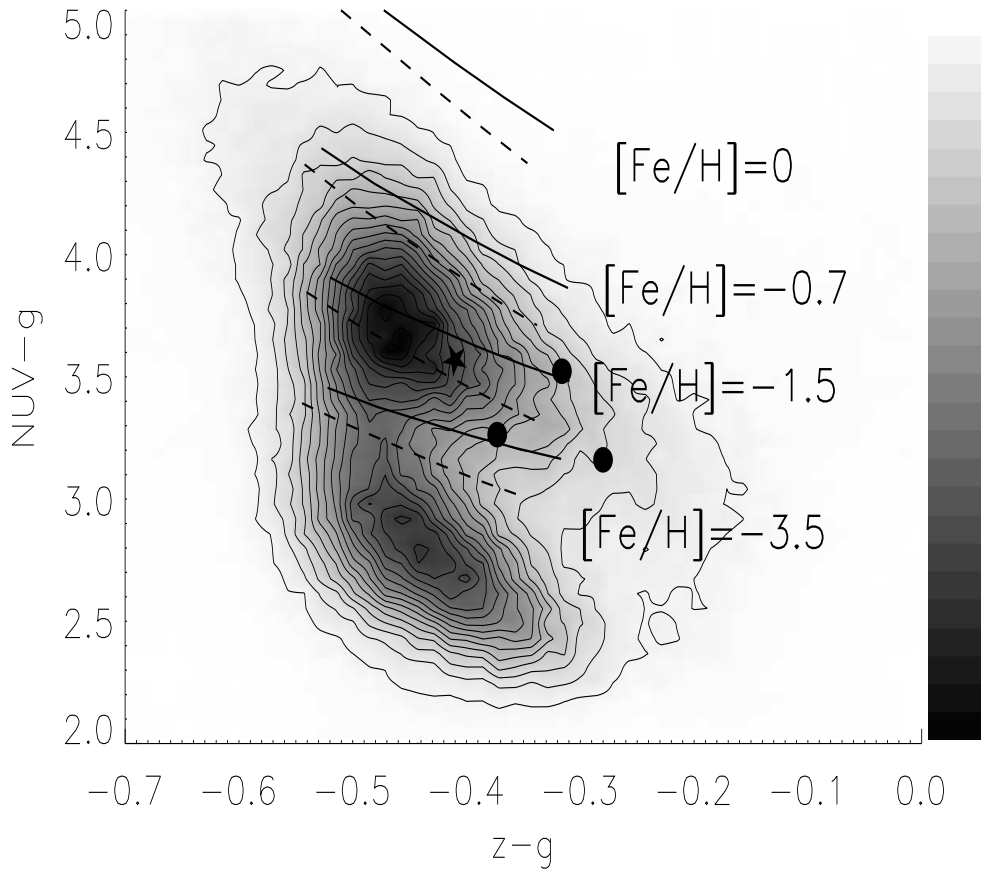}
   \caption{A stellar density map in the plane of the $NUV-g$ and $z-g$ colors for a sample of 200,000
   stars which are photometrically similar to the flux standard BD+17\,4708 (star symbol). Two accumulations of
   sources are present. The first one is centered at $z-g=-0.47$
and $NUV-g=3.7$. The second cluster is at a
similar $z-g$ but a lower $NUV-g\simeq 2.7$. Also, the calculated fluxes for a similar subgiant to BD+17\,4708 are
shown with solid lines for [Fe/H]=0, [Fe/H]$=-0.7$,[Fe/H]=$-1.5$, and [Fe/H]$=-3.5$ metallicity population. The
same models for a dwarf star are illustrated with dashed lines.}
      \label{f1_nuv_g}
 \end{figure*}

Making use of the tables of cross-matched sources between Galaxy Evolution Explorer
({\sc GALEX}) and {\sc SDSS} available at the Multimission Archive at the Space
Telescope ({\sc MAST}), we have endeavoured to identify stars that exhibit colors
of extremely metal-poor stars. Figure~\ref{f1_nuv_g}, a combination of {\sc GALEX} and {\sc SDSS} photometry, i.e., a
stellar density map in the plane of the $NUV\footnote{Near UV (NUV, 1770-2730 \AA\ )}-g$
and $z-g$ colors for a sample of some 200,000 stars, reveals two accumulations of
sources: one centered at $z-g=-0.47$ and $NUV-g=3.7$ and associated with moderately
metal-poor F and G main sequence and subgiant stars. The second cluster at a similar
$z-g$ but at $NUV-g\simeq 2.7$ is populated by  white dwarfs. We have used the $g$, $z$ and
$NUV$ colors of BD\,$+17$\,4708 to set the zero point of the $z-g$ and
$NUV-g$ scales for model fluxes from Kurucz's 
model atmospheres\footnote{http://kurucz.harvard.edu} (Kurucz 1993).
For the photometry of BD\,+17\,4708, the fluxes are taken from Bohlin \& Gilliland (2004); this star is
indicated with a star symbol in Fig.~\ref{f1_nuv_g}. After this calibration, 
the calculated colours for a similar subgiant in the range of
effective temperature between 5900--6400 K with
metallicities typical of the thin disk ([Fe/H]=0), the thick
disk ([Fe/H]$=-0.7$), halo ([Fe/H]=$-1.5$), and ultra-low
metallicity ([Fe/H]$=-4.5$) population, are shown in Fig.~\ref{f1_nuv_g} with 
solid lines: $NUV-g$ decreases as metallicity decreases.  
The same models for a dwarf star ($logg=5$)
correspond to the dashed lines in the figure.

Figure~\ref{f1_nuv_g} is a resource for extracting candidate very metal-poor stars for
detailed abundance analysis. A hasty search of the {\sc MAST} database suggested three
bright stars (Table 1) as candidates for high-resolution optical spectroscopy. Closer
inspection of the database, however, revealed that {\sc SDSS} photometry of the three stars is
unreliable; there are saturation flags with their entries. Thus, the stars are located
incorrectly in Fig.~\ref{f1_nuv_g}, as our abundance analysis demonstrates.

  \begin{table}[!ht]
     \caption[]{Stellar parameters for the three SDSS stars.}
        \label{stel_param}
    $$
        \begin{array}{lllll}
           \hline
           \hline
 Name & T_{eff}  &\log\,g   & \xi &[Fe/H] \\
      & (K)      & (dex)    & km s^{-1}&   \\
           \hline
J100921& 5820\pm125& 3.86\pm0.20& 1.08\pm0.5& -1.30\pm0.13\\ 
J015717& 6250\pm125& 3.70\pm0.20& 4.04\pm0.5& -0.94\pm0.19\\
J171422& 6320\pm125& 4.12\pm0.30& 1.47\pm0.5& -0.80\pm0.15\\ 
\hline				
\hline
        \end{array}
    $$
\end{table}

\section{Spectroscopy}

High-resolution spectra were obtained for the three stars listed in Table 1. One of them,
SDSS J100921.40+375233.9 (which we tagged as SDSS J100921) is a {\it Hipparcos} star, HIP
49750, with a large proper motion. This star is listed in the {\it New Luyten Catalogue of
Stars with Proper Motions Larger than Two Tenths of an Arcsecond} (Luyten 1979b) as NLTT
23519.  The star, SDSS J015717.04+135535.9 (or {\sc SDSS} J015717 in this paper), is
included  in the {\sc 2MASS} catalog ({\sc 2MASS} J01571705+1355360).  The third star, {\sc
SDSS} J171422.43+283657.2 (hereafter {\sc SDSS} J171422) is designated as TYC 2073-1105-1,
and it is also included in the {\sc 2MASS} point source catalog.

 \begin{figure*}[!ht]
 \centering
 \includegraphics*[width=14cm,height=13cm,angle=0]{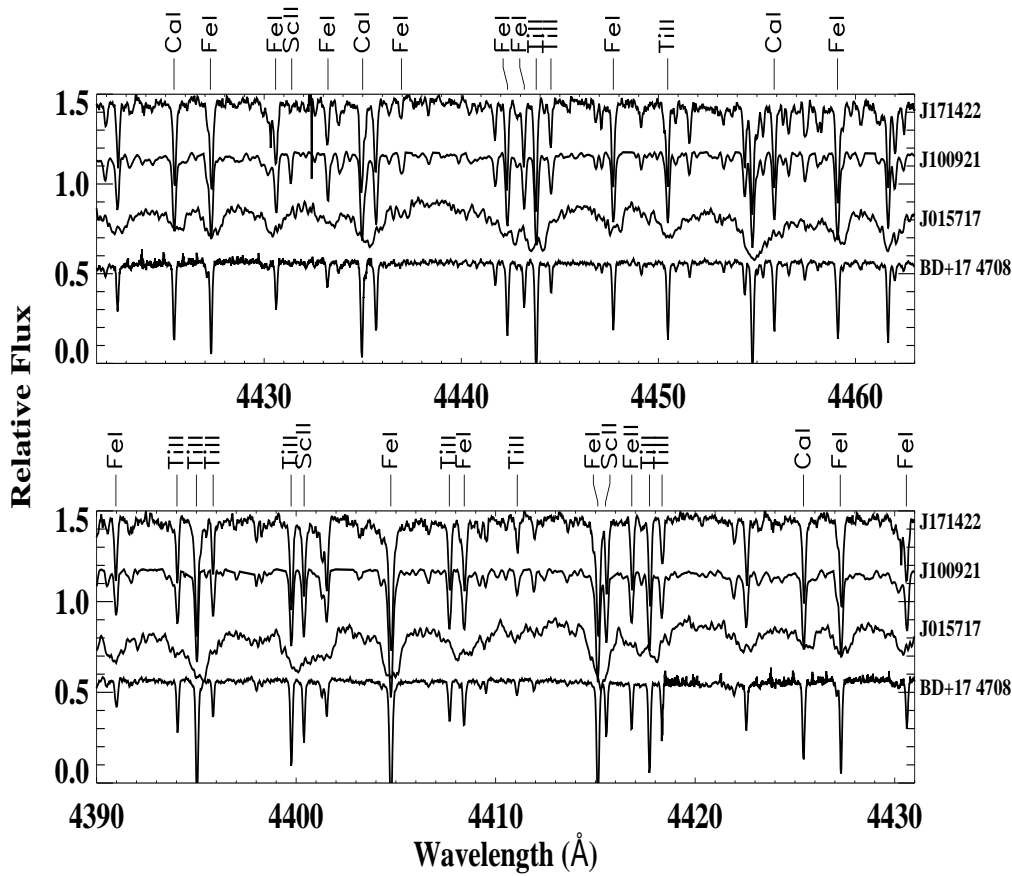}
   \caption{The velocity corrected spectra of SDSS J171422.43+283657.2,
   J100921.40+375233.9, J015717.04+135535.9, and BD\,+17\,4708 over the wavelength regions between
   4390-4431 \AA\ (lower panel) and 4422-4463 \AA\
   (upper panel). Selected lines are identified.}
      \label{f_spectrum_part}
 \end{figure*}

\noindent The spectroscopic data for SDSS SDSS\,J100921 were obtained on 2009 March 15 (two frames) and for J171422 on May 9
(three frames) and for J015717 on November 21 (four frames) 
and December 21 (six frames) at the McDonald observatory with
the 2.7 meter Harlan J. Smith reflector with the CCD-equipped Tull cross-dispersed \'{e}chelle spectrograph
(Tull et al. 1995). The spectra have a FWHM resolving power of 
$\lambda/\delta\lambda \simeq 60,000$ 
with full spectral coverage
from 3600 to 5300 \AA\ , and substantial but incomplete coverage from 5300 to
10\,200 \AA\ . 

\noindent Spectra of 30 minutes each on each night were combined to remove cosmic-rays hits and
to improve the signal-to-noise ratio (S/N). Wavelength calibration was performed using a Th-Ar
hollow cathode lamp for which the spectra were acquired before and after each of the sets of
program star exposures. 

\noindent Observations were reduced using the echelle reduction package in {\sc
IRAF}.\footnote{IRAF is distributed by the National Optical Astronomy Observatories, which is
operated by the Association of Universities for Research in Astronomy (AURA), Inc., under
cooperative agreement with the National Science Foundation.}  The bias level in the overscan area
was modeled with a polynomial and subtracted. In order to correct for pixel-to-pixel sensitivity
variations, a flatfield was used. Scattered light was modeled and removed from the spectrum.
The spectra were optimally extracted. The individual orders were cosmic-ray cleaned up, continuum
normalized, and merged into a continuous spectrum with bespoke echelle reduction software in {\sc
IDL} (\c{S}ahin 2008). The reduced spectra were transferred to the {\sc STARLINK} spectrum analysis
program {\sc DIPSO} (Howarth et al. 1998) for further analysis (e.g. for equivalent width
measurement). A section of the final spectrum is shown in Figure~\ref{f_spectrum_part}. 

\noindent We measured the radial velocity of our targets using cross-correlation
against the spectrum of Arcturus (Hinkle et al. 2000). The spectral range used for the cross-correlation was 4800-5300
\AA\ .\footnote{In the spectra of the program stars, this region is dominated by transitions of
neutral iron, titanium, and nickel.} As a check, we also derived radial velocities by measuring the central wavelength of several
unblended Fe\,{\sc i} lines. These central wavelengths were compared to laboratory values (Nave et al. 1994). The cross-correlation method and the Fe\,{\sc i} method agree with one another within 0.3 km
s$^{-1}$. To correct the small instrumental offsets\footnote{It is 0.2 km s$^{-1}$ for SDSS
J100921, 1.1 km s$^{-1}$ for SDSS J171422, and 0.4 km s$^{-1}$ for SDSS J015717.} between the
spectrum of the star and the hollow-cathode lamp, we use H$_2$O and O$_2$ telluric
lines. The heliocentric corrections were computed with the routine {\sc heliocentric}
in which the routine {\sc baryvel} is called (see Stumpff 1980) from the {\sc IDL} Astronomy User's Library\footnote{\it
http://idlastro.gsfc.nasa.gov}, and applied. For the SDSS stars J100921 and J171422, we determined the heliocentric radial
velocities V$_{\odot}$=$-$59.3$\pm$0.5 km s$^{-1}$ and V$_{\odot}$=26.6$\pm$1.0 km s$^{-1}$,
respectively. For the SDSS star J015717, the heliocentric radial velocities measured on 2009
November 21 and December 21 are -6.6$\pm$1.0 km s$^{-1}$  and -7.9$\pm$0.9 km s$^{-1}$
respectively.

The model atmosphere parameters found for J100921, J015717, and J171422 are
presented in Table 1.

\subsection{ABUNDANCE ANALYSIS - THE MODEL ATMOSPHERES AND STELLAR PARAMETERS}

\noindent Chemical abundances were calculated using the current version of the local
thermodynamic equilibrium (LTE) stellar line analysis program {\sc MOOG} (Sneden
1973). We use a grid of {\sc ATLAS9} model
atmospheres (ODFNEW models) (Castelli \& Kurucz 2003). The models are line-blanketed
plane-parallel atmospheres in LTE and hydrostatic equilibrium with flux (radiative
plus convective) conservation, computed with a constant microturbulent velocity
($\xi=$ 2 km s$^{-1}$). The models are linearly interpolated for the appropriate
values of the spectroscopically determined atmospheric parameters. 

The original sources for the transition probabilities of the Fe\,{\sc i} lines are listed by Lambert et al.
(1996). The $gf$ values for Fe\,{\sc ii} lines are taken from Mel\'endez et al. (2006). These choices for the source of $gf$-values for Fe\,{\sc i} and
Fe\,{\sc ii} lines allow one to make a direct comparison between this study and Ram\'{\i}rez et al. (2006), who
studied BD\,+17\,4708. The revised Fe\,{\sc ii}
line list by Mel\'endez et al. (2006) takes into account new laboratory measurements by Schnabel et al. (2004) and
theoretical calculations by R. L. Kurucz.\footnote{\it See http://kurucz.harvard.edu} The mean difference between the $gf$ values for Fe\,{\sc ii} lines in Lambert et al. (1996)
and Mel\'endez et al. (2006) is just 0.03 dex.

The model parameters were determined using only spectroscopic
criteria. We determined the effective temperature of the star from the condition that
Fe abundance should be independent from the lower excitation potential LEP (lower
excitation potential) of clean Fe\,{\sc i} lines.
Solutions for $T_{\rm eff}$ and $\xi$ are not very sensitive to the adopted surface
gravity. The large range in lower level excitation potentials (LEP) and the large
number of lines\footnote{J100921: EP = 0.0--4.8 eV, $N_{Fe\,{\sc i}}=$105, $N_{Fe\,{\sc
ii}}=$6; J171422: EP = 0.0 -- 4.8, $N_{Fe\,{\sc i}}=$111, $N_{Fe\,{\sc ii}}=$5;
J015717: EP = 0.9 -- 4.8, $N_{Fe\,{\sc i}}=$44, $N_{Fe\,{\sc ii}}=$7} ensures a
reliable identification of the temperature. The determination of the microturbulence velocity $ \xi$, was based on the condition that
iron abundance should be independent of the intensity of the Fe\,{\sc i} line. The determination of T$_{\rm eff}$ and $\xi$ is confirmed by other elements with a large number of
detected lines (e.g. Cr and Ni).

Iron is used as the primary element via ionization equilibrium to provide a
Teff -- log g locus which with an independent estimate of Teff is used to obtain an
estimate of log g. Finally,
the metallicity [Fe/H] is refined by requiring that the derived abundance be equal to
that adopted for the construction of the model atmosphere for the final set of $T_{\rm
eff}$, $log\,g$, and $\xi$.

\subsection{The Balmer Lines}

Balmer line profiles offer an alternative method of estimating atmospheric parameters.
For the three SDSS stars, predicted profiles for H$\beta$ lines are computed with {\sc
SYNTHE} (Kurucz \& Avrett 1981).

 \begin{figure*}[!ht]
 \centering
\includegraphics*[width=140mm,height=12cm,angle=0]{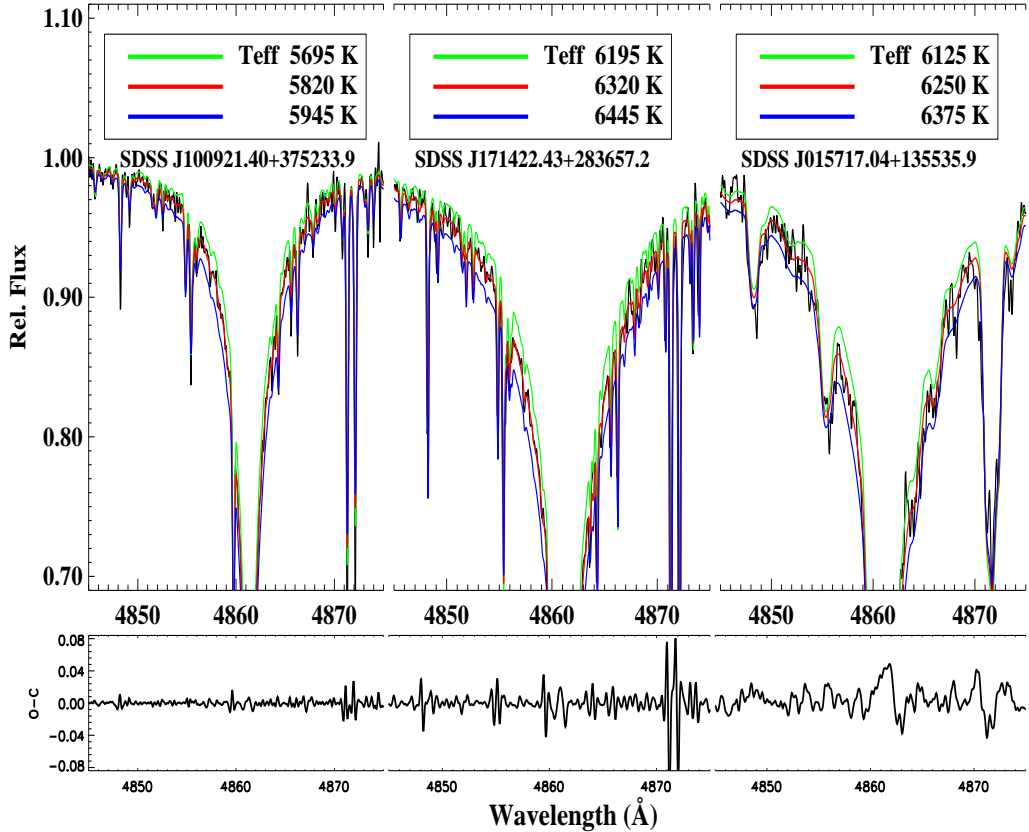}
   \caption{Kurucz model fits for the H$_{\beta}$ line in the spectrum of SDSS J100921 and SDSS
   J171422. The model fits for SDSS J100921 are shown for T$_{\rm eff}=$5820, 6070, and 6320 K and
   log\,g=3.86 and T$_{\rm eff}=$6195, 6320 and 6445 K and log\,g=4.12 for SDSS J171422 and T$_{\rm
   eff}=$ 6125, 6250 and 6375 K and log\,g=3.7 for SDSS J015717. Residuals (observed  minus synthesis)
   of the best fits ((T$_{\rm eff}$, log\,g):(5820, 3.86) for J100921, (6320, 4.12) for J171422, and     
   (6250, 3.70)   for J015717) to the data are also shown.}
      \label{J100921.40+375233.9_587729_balmer}
 \end{figure*}

\noindent Figure~\ref{J100921.40+375233.9_587729_balmer} shows observed and predicted profiles from
synthetic spectrum calculations in the spectra of SDSS J100921 for T$_{\rm eff}$  from 5695 K to
5945 K with log\,g=3.86. For this star, the best-fitting was obtained with a Kurucz model of
T$_{\rm eff}=$5820 K. This temperature is equal to that suggested by the intersection of the Ti
excitation temperature and the Ti, V, and Cr ionization equilibria and very close to the Fe excitation
temperature. In Figure~\ref{J100921.40+375233.9_587729_balmer}, we
also illustrate the observed profiles with best-fitting theoretical line profiles correspond to
effective temperatures of T$_{\rm eff}=$6320 K and 6250 K for log\,g=4.12 and 3.70 and
[Fe/H]=$-$0.75 and $-$0.90 for the stars, J171422 and J015717, respectively. The former temperature for
J171422 is fully consistent with the effective temperature from the excitation of the Fe\,{\sc i}
lines. The latter temperature for the star, J015717 is significantly cooler than
the Fe excitation temperature. These estimates for the temperature are almost independent of
the adopted surface gravities of the program stars. Residuals (observed minus synthesis) of the
best fits to the data are also represented in Figure~\ref{J100921.40+375233.9_587729_balmer}.  

The temperatures obtained from H$_{\beta}$ lines for the SDSS stars, J100921 and J171422 agree well with
the spectroscopic temperatures of the stars. 

\subsection{ABUNDANCE ANALYSIS - ELEMENTS AND LINES}

The lines of neutral and/or singly-ionized atoms were systematically searched for using
lower excitation potentials and $gf$ - values as a guide. The {\sc Revised Multiplet
Table}(RMT) (Moore 1945) is used as starting point in this basic step. Notes on the
adopted $gf$-values with comments on individual elements are presented in Appendix A.
When a reference to solar abundances is necessary in order to convert our abundance of
element X to either of the quantities [X/H] or [X/Fe], Asplund et al. (2009) is adopted.

  \begin{table*}
\tiny
     \caption[]{The Fe lines used in the analysis of SDSS J100921 and J171422 and
     corresponding abundances for models of $T_{\rm eff} = 5820$ K, $\log$ g = 3.9,
     $\xi$ = 1.1 and
    $T_{\rm eff} = 6320$ K, $\log$ g = 4.1, $\xi$ = 1.5, respectively.}
        \label{abund}
    $$
        \begin{array}{l@{}c|@{}r@{}r|@{}c@{}c|cr||l@{}c|@{}r@{}r|@{}c@{}c|cr}
           \hline
     &      & J100921\,\,& J171422 & J100921   & J171422    &      &   &   &  & J100921\,\,& J171422 & J100921 & J171422 &      &   \\
\cline{3-6} \cline{11-14}
 $Species$&\lambda &  $EW$\,\,& $EW$\,\,&\log\epsilon(X)\,\,&\log\epsilon(X)\,\,&$LEP$ &\log(gf) & $Species$&\lambda &  $EW$\,\,& $EW$\,\,&\log\epsilon(X)\,\,&\log\epsilon(X)\,\,&$LEP$ &\log(gf)\\
\cline{3-6} \cline{11-14}

         & ($\AA$)& (m$\AA$) & (m$\AA$)&($dex$) &($dex$)  &($eV$)&      & & ($\AA$)& (m$\AA$) & (m$\AA$)&($dex$) &($dex$)&($eV$)&     \\
           \hline
Fe$\,{\sc i}$ &4494.53&81.0 &   93.0 &6.18&6.76  &2.20 & -1.14 &   Fe$\,{\sc i}$ &5429.70&124.0 &  149.0 &6.40&7.15  &0.96 & -1.88  \\
Fe$\,{\sc i}$ &4630.12&18.0 &   22.0 &6.10&6.61  &2.28 & -2.52 &   Fe$\,{\sc i}$ &5432.95&18.0  &  29.0  &6.56&7.08  &4.45 & -0.94  \\
Fe$\,{\sc i}$ &4745.80&17.0 &   29.0 &6.13&6.74  &3.65 & -1.27 &   Fe$\,{\sc i}$ &5434.52&92.0  &  102.0 &6.18&6.72  &1.01 & -2.12  \\
Fe$\,{\sc i}$ &4800.65&12.0 &   24.0 &6.17&6.81  &4.14 & -1.03 &   Fe$\,{\sc i}$ &5445.02&45.0  &   64.0 &6.15&6.72  &4.39 &  0.04  \\
Fe$\,{\sc i}$ &4918.98&92.0 &  112.0 &6.00&6.62  &2.87 & -0.34 &   Fe$\,{\sc i}$ &5446.92&109.0 &  119.0 &6.24&6.79  &0.99 & -1.91  \\
Fe$\,{\sc i}$ &4924.77&31.0 &   36.0 &6.11&6.59  &2.28 & -2.20 &   Fe$\,{\sc i}$ &5473.90& 23.0 &   31.0 &6.21&6.65  &4.15 & -0.72  \\
Fe$\,{\sc i}$ &4939.69&48.0 &   24.0 &6.24&6.17  &0.86 & -3.34 &   Fe$\,{\sc i}$ &5487.73& 30.0 &   40.0 &6.48&6.92  &4.32 & -0.65  \\
Fe$\,{\sc i}$ &4988.95&27.0 &   46.0 &6.39&7.03  &4.15 & -0.79 &   Fe$\,{\sc i}$ &5497.52& --	&   70.0 & -- &6.77  &1.01 & -2.84  \\
Fe$\,{\sc i}$ &4994.13&55.0 &   54.0 &6.20&6.58  &0.92 & -3.07 &   Fe$\,{\sc i}$ &5501.47& --	&   66.0 & -- &6.82  &0.96 & -3.04  \\
Fe$\,{\sc i}$ &5002.78&20.0 &   30.0 &6.13&6.69  &3.40 & -1.44 &   Fe$\,{\sc i}$ &5506.78& 64.0 &   60.0 &6.19&6.47  &0.99 & -2.80  \\
Fe$\,{\sc i}$ &5012.07&81.0 &   84.0 &6.36&6.76  &0.86 & -2.64 &   Fe$\,{\sc i}$ &5522.44& --	&   11.0 &--  &6.79  &4.21 & -1.40  \\
Fe$\,{\sc i}$ &5014.93&48.0 &   57.0 &6.13&6.53  &3.94 & -0.27 &   Fe$\,{\sc i}$ &5543.94& 9.0  &   18.0 &6.08&6.69  &4.22 & -1.04  \\
Fe$\,{\sc i}$ &5022.21&37.0 &   52.0 &6.18&6.72  &3.98 & -0.49 &   Fe$\,{\sc i}$ &5567.38&10.0  &   16.0 &6.10&6.72  &2.61 & -2.56  \\
Fe$\,{\sc i}$ &5044.21&15.0 &   17.0 &6.03&6.46  &2.85 & -2.03 &   Fe$\,{\sc i}$ &5569.60& 59.0 &   79.0 &6.01&6.65  &3.42 & -0.49  \\
Fe$\,{\sc i}$ &5049.81&77.0 &   82.0 &6.32&6.69  &2.28 & -1.35 &   Fe$\,{\sc i}$ &5572.83& 79.0 &  107.0 &6.15&6.86  &3.40 & -0.28  \\
Fe$\,{\sc i}$ &5051.63&68.0 &   77.0 &6.24&6.79  &0.92 & -2.79 &   Fe$\,{\sc i}$ &5576.08& 44.0 &   57.0 &6.10&6.63  &3.43 & -0.85  \\
Fe$\,{\sc i}$ &5074.73&52.0 &   62.0 &6.35&6.76  &4.22 & -0.16 &   Fe$\,{\sc i}$ &5615.62& 99.0 &  115.0 &6.07&6.59  &3.33 &  0.05  \\
Fe$\,{\sc i}$ &5079.74&59.0 &   44.0 &6.52&6.59  &0.99 & -3.22 &   Fe$\,{\sc i}$ &5633.90& 17.0 &   23.0 &6.22&6.58  &4.99 & -0.12  \\
Fe$\,{\sc i}$ &5083.35&61.0 &   58.0 &6.22&6.53  &0.96 & -2.91 &   Fe$\,{\sc i}$ &5638.26& 18.0 &   27.0 &6.17&6.66  &4.22 & -0.77  \\
Fe$\,{\sc i}$ &5090.76&37.0 &   35.0 &6.27&6.47  &4.26 & -0.36 &   Fe$\,{\sc i}$ &5701.54& 28.0 &   30.0 &6.29&6.70  &2.56 & -2.22  \\
Fe$\,{\sc i}$ &5110.41&77.0 &   75.0 &6.54&6.87  &0.00 & -3.76 &   Fe$\,{\sc i}$ &5717.81& 12.0 &   21.0 &6.22&6.78  &4.28 & -0.98  \\
Fe$\,{\sc i}$ &5121.62&18.0 &   28.0 &6.19&6.70  &4.28 & -0.72 &   Fe$\,{\sc i}$ &5731.75& 14.0 &   14.0 &6.45&6.70  &4.26 & -1.15  \\
Fe$\,{\sc i}$ &5123.72&53.0 &   54.0 &6.23&6.66  &1.01 & -3.07 &   Fe$\,{\sc i}$ &5762.96& 36.0 &   60.0 &6.25&6.98  &4.21 & -0.41  \\
Fe$\,{\sc i}$ &5125.11&54.0 &   68.0 &6.31&6.78  &4.22 & -0.08 &   Fe$\,{\sc i}$ &5930.18& 33.0 &   39.0 &6.35&6.68  &4.65 & -0.17  \\
Fe$\,{\sc i}$ &5127.36&43.0 &   41.0 &6.14&6.55  &0.92 & -3.31 &   Fe$\,{\sc i}$ &5934.65& 17.0 &   25.0 &6.14&6.64  &3.93 & -1.07  \\
Fe$\,{\sc i}$ &5150.84&46.0 &   49.0 &5.97&6.47  &0.99 & -3.00 &   Fe$\,{\sc i}$ &6003.01& 21.0 &   25.0 &6.21&6.59  &3.88 & -1.06  \\
Fe$\,{\sc i}$ &5151.91&37.0 &   42.0 &6.11&6.67  &1.01 & -3.32 &   Fe$\,{\sc i}$ &6027.05& 15.0 &   23.0 &6.24&6.75  &4.08 & -1.09  \\
Fe$\,{\sc i}$ &5159.07&19.0 &   20.0 &6.15&6.43  &4.28 & -0.65 &   Fe$\,{\sc i}$ &6056.00& 18.0 &   24.0 &6.27&6.66  &4.73 & -0.40  \\
Fe$\,{\sc i}$ &5162.25&61.0 &   71.0 &6.26&6.69  &4.18 &  0.08 &   Fe$\,{\sc i}$ &6065.49& 52.0 &   60.0 &6.18&6.64  &2.61 & -1.53  \\
Fe$\,{\sc i}$ &5166.28&53.0 &   62.0 &6.35&7.02  &0.00 & -4.20 &   Fe$\,{\sc i}$ &6136.58& 64.0 &   77.0 &6.17&6.75  &2.45 & -1.40  \\
Fe$\,{\sc i}$ &5171.60&83.0 &  102.0 &6.13&6.82  &1.49 & -1.78 &   Fe$\,{\sc i}$ &6137.65& 61.0 &   71.0 &6.23&6.74  &2.59 & -1.40  \\
Fe$\,{\sc i}$ &5194.94&71.0 &   78.0 &6.22&6.68  &1.56 & -2.08 &   Fe$\,{\sc i}$ &6157.71& 13.0 &   17.0 &6.18&6.59  &4.08 & -1.11  \\
Fe$\,{\sc i}$ &5198.70&39.0 &   40.0 &6.16&6.55  &2.22 & -2.14 &   Fe$\,{\sc i}$ &6170.51& 18.0 &   26.0 &6.31&6.74  &4.79 & -0.38  \\
Fe$\,{\sc i}$ &5216.27&70.0 &   73.0 &6.29&6.67  &1.61 & -2.14 &   Fe$\,{\sc i}$ &6173.31& 11.0 &   18.0 &6.06&6.73  &2.22 & -2.88  \\
Fe$\,{\sc i}$ &5227.17&113.0&  135.0 &6.17&6.86  &1.56 & -1.23 &   Fe$\,{\sc i}$ &6191.55& --	&   72.0 &--  &6.61  &2.43 & -1.42  \\
Fe$\,{\sc i}$ &5228.38&13.0 &   19.0 &6.43&6.88  &4.22 & -1.19 &   Fe$\,{\sc i}$ &6213.43& 25.0 &   23.0 &6.12&6.47  &2.22 & -2.48  \\
Fe$\,{\sc i}$ &5232.92&113.0&  126.0 &6.07&6.57  &2.94 & -0.08 &   Fe$\,{\sc i}$ &6232.64& 25.0 &   28.0 &6.26&6.62  &3.65 & -1.22  \\
Fe$\,{\sc i}$ &5242.48&32.0 &   39.0 &6.20&6.62  &3.63 & -0.97 &   Fe$\,{\sc i}$ &6252.52& 61.0 &   60.0 &6.37&6.66  &2.40 & -1.72  \\
Fe$\,{\sc i}$ &5250.19&12.0 &   15.0 &6.11&6.80  &0.12 & -4.92 &   Fe$\,{\sc i}$ &6265.13& 28.0 &   37.0 &6.22&6.81  &2.18 & -2.55  \\
Fe$\,{\sc i}$ &5253.46&21.0 &   27.0 &6.17&6.64  &3.28 & -1.57 &   Fe$\,{\sc i}$ &6322.69& 23.0 &   24.0 &6.38&6.77  &2.59 & -2.43  \\
Fe$\,{\sc i}$ &5288.51&12.0 &   18.0 &6.20&6.71  &3.69 & -1.51 &   Fe$\,{\sc i}$ &6344.15& 8.0  &   15.0 &6.14&6.86  &2.43 & -2.92  \\
Fe$\,{\sc i}$ &5302.29&66.0 &   69.0 &6.30&6.62  &3.28 & -0.74 &   Fe$\,{\sc i}$ &6380.71&  --  &   13.0 &--  &6.75  &4.19 & -1.32  \\
Fe$\,{\sc i}$ &5307.36&27.0 &   34.0 &6.11&6.70  &1.61 & -2.98 &   Fe$\,{\sc i}$ &6393.56& 58.0 &   70.0 &6.03&6.60  &2.43 & -1.43  \\
Fe$\,{\sc i}$ &5315.07& --  &    8.0 & -- &6.78  &4.37 & -1.40 &   Fe$\,{\sc i}$ &6411.63& 53.0 &   65.0 &6.26&6.75  &3.65 & -0.66  \\
Fe$\,{\sc i}$ &5322.04&9.0  &   18.0 &6.06&6.83  &2.28 & -2.89 &   Fe$\,{\sc i}$ &6419.95& --	&   36.0 &--  &6.79  &4.73 & -0.27  \\
Fe$\,{\sc i}$ &5324.17&95.0 &  105.0 &6.08&6.53  &3.21 & -0.10 &   Fe$\,{\sc i}$ &6421.34& --	&   59.0 &--  &6.79  &2.28 & -2.01  \\
Fe$\,{\sc i}$ &5332.90&32.0 &   37.0 &5.98&6.52  &1.56 & -2.78 &   Fe$\,{\sc i}$ &6592.89& 50.0 &   55.0 &6.17&6.57  &2.73 & -1.47  \\
Fe$\,{\sc i}$ &5341.02&88.0 &   90.0 &6.50&6.84  &1.61 & -1.95 &   Fe$\,{\sc i}$ &6593.85& 22.0 &   29.0 &6.18&6.73  &2.44 & -2.42  \\
Fe$\,{\sc i}$ &5353.38&27.0 &   40.0 &6.22&6.75  &4.10 & -0.68 &   Fe$\,{\sc i}$ &6597.51& --	&   19.0 &--  &7.09  &4.79 & -0.92  \\
Fe$\,{\sc i}$ &5364.85&51.0 &   68.0 &6.13&6.64  &4.45 &  0.23 &   Fe$\,{\sc i}$ &6609.11& 12.0 &   20.0 &6.23&6.89  &2.56 & -2.69  \\
Fe$\,{\sc i}$ &5367.44&57.0 &   76.0 &6.01&6.61  &4.42 &  0.44 &   Fe$\,{\sc i}$ &6750.15& 22.0 &   23.0 &6.36&6.77  &2.42 & -2.62  \\
Fe$\,{\sc i}$ &5369.93&61.0 &   75.0 &5.95&6.46  &4.37 &  0.54 &   Fe$\,{\sc i}$ &6841.34& 10.0 &   20.0 &6.14&6.73  &4.61 & -0.71  \\
Fe$\,{\sc i}$ &5371.49&120.0&  153.0 &6.12&6.96  &0.96 & -1.65 &   Fe$\,{\sc i}$ &6855.16& 16.0 &   29.0 &6.37&6.95  &4.56 & -0.74  \\

\hline
        \end{array}
    $$
\end{table*}

  \begin{table*}
\tiny
     \caption[]{The Fe lines used in the analysis of SDSS J100921 and J171422 and
     corresponding abundances for models of $T_{\rm eff} = 5820$ K, $\log$ g = 3.9,
     $\xi$ = 1.1 and
    $T_{\rm eff} = 6320$ K, $\log$ g = 4.1, $\xi$ = 1.5, respectively.}
        \label{abund}
    $$
        \begin{array}{l@{}c|@{}r@{}r|@{}c@{}c|cr||l@{}c|@{}r@{}r|@{}c@{}c|cr}
           \hline
     &      & J100921\,\,& J171422 & J100921   & J171422    &      &   &   &  & J100921\,\,& J171422 & J100921 & J171422 &      &   \\
\cline{3-6} \cline{11-14}
 $Species$&\lambda &  $EW$\,\,& $EW$\,\,&\log\epsilon(X)\,\,&\log\epsilon(X)\,\,&$LEP$ &\log(gf) & $Species$&\lambda &  $EW$\,\,& $EW$\,\,&\log\epsilon(X)\,\,&\log\epsilon(X)\,\,&$LEP$ &\log(gf)\\
\cline{3-6} \cline{11-14}

         & ($\AA$)& (m$\AA$) & (m$\AA$)&($dex$) &($dex$)  &($eV$)&      & & ($\AA$)& (m$\AA$) & (m$\AA$)&($dex$) &($dex$)&($eV$)&     \\
           \hline

Fe$\,{\sc i}$ &5373.71&12.0 &   19.0 &6.17&6.65  &4.47 & -0.74 &   Fe$\,{\sc i}$ &7130.89& 22.0 &   30.0 &6.16&6.61  &4.22 & -0.70  \\
Fe$\,{\sc i}$ &5383.36&70.0 &   94.0 &5.97&6.57  &4.31 &  0.65 &   Fe$\,{\sc ii}$&4620.52& 24.0 &   35.0 &6.21&6.60  &2.83 & -3.21  \\
Fe$\,{\sc i}$ &5389.47&26.0 &   36.0 &6.06&6.51  &4.42 & -0.25 &   Fe$\,{\sc ii}$&4629.34&  57.0&   76.0 &6.08&6.55  &2.81 & -2.28  \\
Fe$\,{\sc i}$ &5393.16&58.0 &   78.0 &6.08&6.73  &3.24 & -0.72 &   Fe$\,{\sc ii}$&5197.58&  48.0&   68.0 &6.17&6.64  &3.23 & -2.22  \\
Fe$\,{\sc i}$ &5397.13&101.0&  123.0 &6.13&6.88  &0.92 & -1.99 &   Fe$\,{\sc ii}$&5234.62&  54.0&   75.0 &6.27&6.75  &3.22 & -2.18  \\
Fe$\,{\sc i}$ &5405.76&106.0&  109.0 &6.14&6.56  &0.99 & -1.85 &   Fe$\,{\sc ii}$&5264.81&  16.0&   28.0 &6.25&6.70  &3.23 & -3.13  \\
Fe$\,{\sc i}$ &5410.89&60.0 &   68.0 &6.15&6.49  &4.47 &  0.40 &   Fe$\,{\sc ii}$&6516.08&  20.0&   --   &6.18& --   &2.89 & -3.31  \\
\hline
        \end{array}
    $$
\end{table*}


  \begin{table*}
\tiny
     \caption[]{The lines used in the analysis of SDSS J100921 and SDSS J171422 and
     corresponding abundances for models of $T_{\rm eff} = 5820$ K, $\log$ g = 3.9,
     $\xi$ = 1.1 and
    $T_{\rm eff} = 6320$ K, $\log$ g = 4.1, $\xi$ = 1.5, respectively.}
        \label{abund}
    $$
        \begin{array}{l@{}c|@{}r@{}r|@{}c@{}c|cr||l@{}c|@{}r@{}r|@{}c@{}c|cr}
           \hline
         &        & J100921\,\,& J171422 & J100921  & J171422  &   &   &   &   & J100921\,\,& J171422 & J100921  & J171422  &      &       \\
\cline{3-6} \cline{11-14}
 $Species$&\lambda &  $EW$\,\,& $EW$\,\,&\log\epsilon(X)\,\,&\log\epsilon(X)\,\,&$LEP$ &\log(gf) & $Species$&\lambda &  $EW$\,\,& $EW$\,\,&\log\epsilon(X)\,\,&\log\epsilon(X)\,\,&$LEP$ &\log(gf)\\
\cline{3-6} \cline{11-14}
          & ($\AA$)& (m$\AA$) & (m$\AA$)&($dex$)               &($dex$)&($eV$)&    & ($\AA$)& (m$\AA$) & (m$\AA$)&($dex$)               &($dex$)&($eV$)&    \\
           \hline
C$\,{\sc i}$  &8335.15&  SS  & SS   &7.76&7.81  &7.69 & -0.44 & Ti$\,{\sc ii}$&4417.72& 79.0 & 84.0 &4.27 &4.46 &1.17 & -1.43 \\
C$\,{\sc i}$  &9061.44&  --  & SS   & -- &8.11  &7.48 & -0.35 & Ti$\,{\sc ii}$&4418.33& 44.0 & 49.0 &3.83 &4.14 &1.24 & -1.82 \\
C$\,{\sc i}$  &9062.49&  SS  & SS   &7.58&8.06  &7.48 & -0.46 & Ti$\,{\sc ii}$&4443.76& 103.0&116.0 &4.03 &4.39 &1.08 & -0.71 \\
C$\,{\sc i}$  &9078.29&  41.0& --   &7.41& --	&7.48 & -0.58 & Ti$\,{\sc ii}$&4444.51& 40.0 & 41.0 &3.83 &4.09 &1.12 & -2.03 \\
O$\,{\sc i}$  &7771.94&  58.0& 93.0 &8.33&8.58  &9.15 &  0.37 & Ti$\,{\sc ii}$&4450.44& 71.0 & 90.0 &3.98 &4.54 &1.08 & -1.45 \\
O$\,{\sc i}$  &7774.17&  44.0& 83.0 &8.25&8.59  &9.15 &  0.22 & Ti$\,{\sc ii}$&4468.49& 115.0&115.0 &4.20 &4.31 &1.13 & -0.62 \\
O$\,{\sc i}$  &7775.39&  30.0& 59.0 &8.19&8.45  &9.15 &  0.00 & Ti$\,{\sc ii}$&4493.46& 16.0 & 16.0 &3.88 &4.18 &1.08 & -2.74 \\
O$\,{\sc i}$  &8446.25&  SS  & SS   &8.33&8.38  &9.52 & -0.46 & Ti$\,{\sc ii}$&4501.23& 106.0&116.0 &4.16 &4.45 &1.12 & -0.75 \\
O$\,{\sc i}$  &8446.36&  SS  & SS   &8.33&8.38  &9.52 &  0.24 & Ti$\,{\sc ii}$&4544.03& 20.0 & 20.0 &3.83 &4.11 &1.24 & -2.41 \\
O$\,{\sc i}$  &8446.76&  SS  & SS   &8.33&8.38  &9.52 &  0.01 & Ti$\,{\sc ii}$&4563.72& 99.0 &104.0 &4.30 &4.47 &1.22 & -0.96 \\
Na$\,{\sc i}$ &5682.63&  --  & 38.0 & -- & 5.64 &2.10 & -0.71 & Ti$\,{\sc ii}$&4571.94& 111.0&119.0 &4.43 &4.66 &1.57 & -0.52 \\
Na$\,{\sc i}$ &5688.21&  SS  & SS   &5.20& 5.65 &2.10 & -0.45 & Ti$\,{\sc ii}$&5154.06& 44.0 & 42.0 &4.20 &4.36 &1.57 & -1.92 \\
Na$\,{\sc i}$ &8183.26&  --  & 199.0& -- & 5.98 &2.10 &  0.24 & Ti$\,{\sc ii}$&5336.76& 48.0 & 50.0 &4.07 &4.29 &1.58 & -1.70 \\
Na$\,{\sc i}$ &8194.79&  --  & SS   & -- & 6.13 &2.10 & -0.46 & Ti$\,{\sc ii}$&5381.01& 37.0 & 38.0 &4.19 &4.43 &1.57 & -2.08 \\
Mg$\,{\sc i}$ &4571.10&  52.0&36.0  &6.54& 6.78 &0.00 & -5.62 & Ti$\,{\sc ii}$&5418.74& 31.0 & 24.0 &3.98 &4.05 &1.58 & -1.99 \\
Mg$\,{\sc i}$ &5711.10&  42.0&54.0  &6.58&6.98  &4.35 & -1.67 & V$\,{\sc i}$  &4444.21& SS   &  SS  &2.59 &3.47 &0.27 & -0.72 \\
Al$\,{\sc i}$ &3944.02& SS   & SS   &5.12&5.57  &0.00 & -0.64 & V$\,{\sc ii}$ &3903.22& --   & 49.0 &--   &3.43 &1.48 & -0.89 \\
Al$\,{\sc i}$ &3961.52& SS   & SS   &5.02&5.32  &0.01 & -0.34 & V$\,{\sc ii}$ &3951.94& 37.0 & 52.0 &2.80 &3.33 &1.48 & -0.74 \\
Si$\,{\sc i}$ &5708.40&  24.0&26.0  &6.52&6.75  &4.95 & -1.37 & V$\,{\sc ii}$ &4005.69& --   & 59.0 & --  &3.51 &1.82 & -0.46 \\
Si$\,{\sc i}$ &5948.54&  23.0&36.0  &6.38&6.83  &5.08 & -1.13 & Cr$\,{\sc i}$ &3991.10& 21.0 & 35.0 &4.02 &4.64 &2.54 &  0.25 \\
Ca$\,{\sc i}$ &4526.93&  33.0&35.0  &5.16&5.43  &2.71 & -0.42 & Cr$\,{\sc i}$ &4274.76& 118.0&117.0 &4.28 &4.67 &0.00 & -0.23 \\
Ca$\,{\sc i}$ &4578.55&  34.0&33.0  &5.16&5.37  &2.52 & -0.56 & Cr$\,{\sc i}$ &4496.84& 46.0 & 53.0 &4.45 &4.95 &0.94 & -1.15 \\
Ca$\,{\sc i}$ &5512.98&  34.0&30.0  &5.26&5.38  &2.93 & -0.30 & Cr$\,{\sc i}$ &4545.96& 29.0 & 28.0 &4.27 &4.66 &0.94 & -1.38 \\
Ca$\,{\sc i}$ &5857.45&  67.0&83.0  &5.37&5.82  &2.93 &  0.23 & Cr$\,{\sc i}$ &4591.42& 18.0 & 22.0 &4.38 &4.92 &0.97 & -1.76 \\
Ca$\,{\sc i}$ &6162.16& 115.0&123.0 &5.30&5.71  &1.90 & -0.08 & Cr$\,{\sc i}$ &4600.75& 30.0 & 30.0 &4.23 &4.64 &1.00 & -1.26 \\
Ca$\,{\sc i}$ &6166.44&  17.0&21.0  &5.01&5.39  &2.52 & -0.90 & Cr$\,{\sc i}$ &4616.09& 41.0 & 39.0 &4.38 &4.72 &0.98 & -1.18 \\
Ca$\,{\sc i}$ &6169.04&  31.0&43.0  &5.01&5.49  &2.52 & -0.54 & Cr$\,{\sc i}$ &4622.41& --   & 11.0 &--   &5.10 &3.55 & -0.04 \\
Ca$\,{\sc i}$ &6169.56&  54.0&60.0  &5.18&5.52  &2.53 & -0.27 & Cr$\,{\sc i}$ &4626.14& 33.0 & 34.0 &4.33 &4.75 &0.97 & -1.32 \\
Ca$\,{\sc i}$ &6439.07&  97.0&107.0 &5.33&5.67  &2.53 &  0.47 & Cr$\,{\sc i}$ &4646.15& 53.0 & 56.0 &4.25 &4.64 &1.03 & -0.71 \\
Ca$\,{\sc i}$ &6471.66&  32.0&42.0  &5.11&5.54  &2.53 & -0.59 & Cr$\,{\sc i}$ &4651.30& 20.0 & 25.0 &4.14 &4.70 &0.98 & -1.46 \\
\hline
        \end{array}
    $$
\end{table*}

  \begin{table*}
\tiny
     \caption[]{The lines used in the analysis of SDSS J100921 and SDSS J171422 and
     corresponding abundances for models of $T_{\rm eff} = 5820$ K, $\log$ g = 3.9,
     $\xi$ = 1.1 and
    $T_{\rm eff} = 6320$ K, $\log$ g = 4.1, $\xi$ = 1.5, respectively.}
        \label{abund}
    $$
        \begin{array}{l@{}c|@{}r@{}r|@{}c@{}c|cr||l@{}c|@{}r@{}r|@{}c@{}c|cr}
           \hline
         &        & J100921\,\,& J171422 & J100921  & J171422  &   &   &   &   & J100921\,\,& J171422 & J100921  & J171422  &      &       \\
\cline{3-6} \cline{11-14}
 $Species$&\lambda &  $EW$\,\,& $EW$\,\,&\log\epsilon(X)\,\,&\log\epsilon(X)\,\,&$LEP$ &\log(gf) & $Species$&\lambda &  $EW$\,\,& $EW$\,\,&\log\epsilon(X)\,\,&\log\epsilon(X)\,\,&$LEP$ &\log(gf)\\
\cline{3-6} \cline{11-14}
          & ($\AA$)& (m$\AA$) & (m$\AA$)&($dex$)               &($dex$)&($eV$)&    & ($\AA$)& (m$\AA$) & (m$\AA$)&($dex$)               &($dex$)&($eV$)&    \\
           \hline
Ca$\,{\sc i}$ &6493.78&  86.0&80.0  &5.46&5.52  &2.52 &  0.14 & Cr$\,{\sc i}$ &4652.17& 41.  & 44.0 &4.25 &4.69 &1.00 & -1.03 \\
Ca$\,{\sc i}$ &6499.65&  20.0&40.0  &4.80&5.50  &2.52 & -0.59 & Cr$\,{\sc i}$ &4708.00& 11.0 & 18.0 &4.36 &4.88 &3.17 &  0.11 \\
Sc$\,{\sc ii}$&4246.83& 127.0&126.0 &2.38 &2.59 &0.32 &  0.24 & Cr$\,{\sc i}$ &4718.42& 10.0 & 22.0 &4.34 &5.02 &3.20 &  0.10 \\
Sc$\,{\sc ii}$&4314.08&  --  &102.0 & --  &2.68 &0.62 & -0.10 & Cr$\,{\sc i}$ &4756.11&13.0  & 20.0 &4.40&4.90  &3.10 &  0.09 \\
Sc$\,{\sc ii}$&4670.40& 30.0 & 33.0 &1.98 &2.30 &1.36 & -0.58 & Cr$\,{\sc i}$ &4789.36&18.0  & 19.0 &4.49&4.83  &2.54 & -0.37 \\
Sc$\,{\sc ii}$&5030.98& 37.0 & 52.0 &1.94 &2.46 &1.36 & -0.40 & Cr$\,{\sc i}$ &4922.25&25.0  & 26.0 &4.57&4.85  &3.10 &  0.28 \\
Sc$\,{\sc ii}$&5239.80& 21.0 & 28.0 &2.01 &2.44 &1.46 & -0.77 & Cr$\,{\sc i}$ &5247.56& 20.0 & 22.0 &4.26&4.74  &0.96 & -1.63 \\
Sc$\,{\sc ii}$&5526.78& 43.0 & 59.0 &2.03 &2.52 &1.77 &  0.02 & Cr$\,{\sc i}$ &5296.68& 28.0 & 31.0 &4.27&4.75  &0.98 & -1.41 \\
Sc$\,{\sc ii}$&5640.95& 15.0 & 21.0 &1.95 &2.39 &1.50 & -0.87 & Cr$\,{\sc i}$ &5297.37& 20.0 & 30.0 &4.34&4.86  &2.90 &  0.17 \\
Sc$\,{\sc ii}$&5657.85& 33.0 & 43.0 &2.17 &2.60 &1.51 & -0.60 & Cr$\,{\sc i}$ &5300.74& --   & 10.0 &--  &4.84  &0.98 & -2.13 \\
Sc$\,{\sc ii}$&6245.58& 11.0 & 16.0 &1.95 &2.41 &1.51 & -1.05 & Cr$\,{\sc i}$ &5329.14& 9.0  & 16.0 &4.16&4.74  &2.91 & -0.06 \\
Sc$\,{\sc ii}$&6604.57& 14.0 & 18.0 &2.16 &2.58 &1.36 & -1.31 & Cr$\,{\sc i}$ &5345.78& 53.0 & 56.0 &4.42&4.83  &1.00 & -0.98 \\
Ti$\,{\sc i}$ &3998.64& 77.0 & 68.0 &4.14 &4.23 &0.05 & -0.06 & Cr$\,{\sc i}$ &5348.29& 33.0 & 38.0 &4.28&4.79  &1.00 & -1.29 \\
Ti$\,{\sc i}$ &4518.02& 29.0 & 28.0 &3.91 &4.31 &0.83 & -0.32 & Cr$\,{\sc i}$ &5409.77& 55.0 & 67.0 &4.24&4.82  &1.03 & -0.72 \\
Ti$\,{\sc i}$ &4533.24& 65.0 & 71.0 &3.96 &4.40 &0.85 &  0.48 & Cr$\,{\sc ii}$&4554.96& 19.0 & 23.0 &4.63&4.81  &4.07 & -1.37 \\
Ti$\,{\sc i}$ &4534.78& 50.0 & 54.0 &3.79 &4.23 &0.84 &  0.28 & Cr$\,{\sc ii}$&4558.60& 50.0 & 65.0 &4.72&5.03  &4.07 & -0.66 \\
Ti$\,{\sc i}$ &4552.43& 49.0 & 51.0 &4.40 &4.80 &0.84 & -0.34 & Cr$\,{\sc ii}$&4588.18& 28.0 & 54.0 &4.15&4.76  &4.07 & -0.64 \\
Ti$\,{\sc i}$ &4555.45& 17.0 & 23.0 &3.77 &4.37 &0.85 & -0.49 & Cr$\,{\sc ii}$&4592.04& 17.0 & 32.0 &4.41&4.87  &4.07 & -1.22 \\
Ti$\,{\sc i}$ &4722.64& --   &  SS  &--   &4.70 &1.05 & -1.33 & Cr$\,{\sc ii}$&4616.59& 17.0 & 25.0 &4.48&4.78  &4.07 & -1.29 \\
Ti$\,{\sc i}$ &4840.87& 20.0 & 23.0 &3.91 &4.42 &0.90 & -0.51 & Cr$\,{\sc ii}$&4812.31& 7.0  & 16.0 &4.33&4.84  &3.86 & -1.80 \\
Ti$\,{\sc i}$ &4885.02& 8.0  & 23.0 &3.53 &4.43 &1.89 &  0.36 & Cr$\,{\sc ii}$&4848.23& 25.0 & 38.0 &4.35&4.72  &3.86 & -1.13 \\
Ti$\,{\sc i}$ &4981.71& 70.0 & 68.0 &4.02 &4.28 &0.85 &  0.50 & Cr$\,{\sc ii}$&4876.39& 24.0 & 37.0 &4.66&5.03  &3.86 & -1.47 \\
Ti$\,{\sc i}$ &4999.50& 52.0 & 56.0 &3.82 &4.26 &0.83 &  0.25 & Cr$\,{\sc ii}$&5237.31& 24.0 & 37.0 &4.55&4.90  &4.07 & -1.16 \\
Ti$\,{\sc i}$ &5016.15& 16.0 & 17.0 &3.79 &4.26 &0.85 & -0.57 & Cr$\,{\sc ii}$&5308.41& --   & 13.0 &--  &4.91  &4.07 & -1.81 \\
Ti$\,{\sc i}$ &5022.90& 14.0 & 25.0 &3.56 &4.32 &0.83 & -0.43 & Mn$\,{\sc i}$ &3823.51&40.0  & 66.0 &3.74&4.64  & 2.14&  0.06 \\
Ti$\,{\sc i}$ &5036.43& 21.0 & 21.0 &3.82 &4.20 &1.44 &  0.13 & Mn$\,{\sc i}$ &3823.89& --   & SS   & -- &4.54  & 2.16& -0.52 \\
Ti$\,{\sc i}$ &5192.95& 31.0 & 25.0 &3.80 &4.15 &0.02 & -1.01 & Mn$\,{\sc i}$ &4034.45& 117.0&110.0 &4.35&4.65  & 0.00& -0.81 \\
Ti$\,{\sc i}$ &5210.38& 37.0 & 37.0 &3.84 &4.30 &0.05 & -0.88 & Mn$\,{\sc i}$ &4048.73&53.0  & 60.0 &4.25& 4.67 & 2.16& -0.13 \\
Ti$\,{\sc i}$ &6258.65& 14.0 & 20.0 &3.93 &4.51 &1.46 & -0.24 & Mn$\,{\sc i}$ &4055.53&38.0  & 47.0 &3.77& 4.29 & 2.13& -0.07 \\
Ti$\,{\sc ii}$&4316.80& 30.0 & 34.0 &3.91 &4.19 &2.05 & -1.42 & Mn$\,{\sc i}$ &4436.33& 9.0  & 18.0 &3.83&4.50  & 2.91& -0.29 \\
Ti$\,{\sc ii}$&4394.02& 57.0 & 55.0 &4.20 &4.32 &1.22 & -1.89 & Mn$\,{\sc i}$ &4470.14& --   & 14.0 &--  &4.54  & 2.93& -0.44 \\
Ti$\,{\sc ii}$&4395.81& 41.0 & 45.0 &4.12 &4.42 &1.24 & -2.17 & Mn$\,{\sc i}$ &4709.71& 6.0  & 19.0 &3.64&4.55  & 2.88& -0.34 \\
Mn$\,{\sc i}$ &4739.07& --   & 16.0 &--  &4.64  & 2.93& -0.49 & Ni$\,{\sc i}$ &5082.32& 12.0 &16.0  &4.88& 5.31 & 3.66& -0.54  \\
Mn$\,{\sc i}$ &4761.52& --   & 23.0 &--  &4.51  & 2.94& -0.14 & Ni$\,{\sc i}$ &5084.10& 35.0 &40.0  &4.98& 5.34 & 3.68&  0.03  \\
Mn$\,{\sc i}$ &4783.42& 52.0 & 64.0 &4.09&4.62  & 2.29&  0.04 & Ni$\,{\sc i}$ &5115.37& 29.0 &29.0  &5.13& 5.38 & 3.83& -0.11  \\
Mn$\,{\sc i}$ &4823.47& 51.0 & 71.0 &3.98&4.70  & 2.31&  0.14 & Ni$\,{\sc i}$ &5137.06& 31.0 &44.0  &4.98& 5.65 & 1.68& -1.99  \\
Mn$\,{\sc i}$ &6013.49& 7.0  & 12.0 &3.75&4.33  & 3.06& -0.25 & Ni$\,{\sc i}$ &5155.77& 19.0 &25.0  &4.91& 5.32 & 3.90&-0.09   \\
Mn$\,{\sc i}$ &6016.64& --   & 22.0 &--  &4.62  & 3.06& -0.22 & Ni$\,{\sc i}$ &5578.75& 9.0  &11.0  &4.89& 5.43 & 1.68&-2.64   \\
Mn$\,{\sc i}$ &6021.76& 12.0 & 29.0 &3.73&4.54  & 3.06&  0.03 & Ni$\,{\sc i}$ &5754.63& 17.0 &15.0  &5.17& 5.52 & 1.94&-2.34   \\
Co$\,{\sc i}$ &3995.30& 77.0 & 83.0 &3.98& 4.43 & 0.92& -0.22 & Ni$\,{\sc i}$ &6175.36& 6.0  &15.0  &4.92& 5.63 & 4.09&-0.54   \\
Co$\,{\sc i}$ &4121.28& 73.0 & 72.0 &3.98& 4.24 & 0.92& -0.32 & Ni$\,{\sc i}$ &6176.78& 11.0 &17.0  &5.21& 5.69 & 4.09&-0.53   \\
Ni$\,{\sc i}$ &4331.63& 17.0 & 11.0 &4.76& 4.96 & 1.68& -2.10 & Ni$\,{\sc i}$ &6643.60& 29.0 &27.0  &5.17& 5.54 & 1.68&-2.30   \\
Ni$\,{\sc i}$ &4462.42&18.0  & 19.0 &5.01& 5.33 & 3.46& -0.60 & Ni$\,{\sc i}$ &6767.74& 21.0 &27.0  &4.98& 5.54 & 1.83&-2.17   \\
Ni$\,{\sc i}$ &4470.47& 28.0 &28.0  &5.01& 5.29 & 3.40& -0.40 & Ni$\,{\sc i}$ &7122.18& 32.0 &60.0  &4.68& 5.49 & 3.54& 0.05   \\
Ni$\,{\sc i}$ &4604.97& 27.0 &--    &4.95& --	& 3.48& -0.29 & Ni$\,{\sc i}$ &7422.26& 31.0 &38.0  &4.91& 5.33 & 3.63&-0.13   \\
Ni$\,{\sc i}$ &4648.66& 37.0 &40.0  &4.98& 5.31 & 3.42& -0.15 & Ni$\,{\sc i}$ &7788.90& 17.0 &25.0  &5.18& 5.81 & 1.95&-2.42   \\
Ni$\,{\sc i}$ &4686.17& 13.0 &15.0  &4.99& 5.34 & 3.60& -0.64 & Zn$\,{\sc i}$ &4722.14& 28.0 & 32.0 &3.39&3.71  &4.03 &-0.39   \\
Ni$\,{\sc i}$ &4715.75& 25.0 &31.0  &4.99& 5.40 & 3.54& -0.33 & Zn$\,{\sc i}$ &4810.50& 32.0 & 38.0 &3.32&3.66  &4.08 &-0.17   \\
Ni$\,{\sc i}$ &4756.51& 25.0 &28.0  &4.94& 5.29 & 3.48& -0.34 & Sr$\,{\sc ii}$&4077.71& 161.0&178.0 &1.45& 1.99 & 0.00& 0.14   \\
Ni$\,{\sc i}$ &4806.96& 12.0 &18.0  &5.02& 5.51 & 3.68& -0.64 & Sr$\,{\sc ii}$&4215.52& 157.0&171.0 &1.70& 2.22 & 0.00& -0.18  \\
Ni$\,{\sc i}$ &4829.00& 28.0 &38.0  &5.06& 5.54 & 3.54& -0.33 & Y$\,{\sc ii}$ &3950.35& 42.0 &45.0  &0.94& 1.29 &0.10 & -0.49  \\
Ni$\,{\sc i}$ &4831.16& 18.0 &24.0  &4.93& 5.37 & 3.61& -0.41 & Y$\,{\sc ii}$ &4883.69& 28.0 &37.0  &0.89& 1.36 &1.08 &  0.07  \\
Ni$\,{\sc i}$ &4852.57& --   &10.0  &--  & 5.51 & 3.54& -1.07 & Y$\,{\sc ii}$ &4900.12& SS   & SS   &0.84& 1.49 &1.03 & -0.09  \\
Ni$\,{\sc i}$ &4904.40& 30.0 &40.0  &4.94& 5.42 & 3.54& -0.17 & Y$\,{\sc ii}$ &5087.43& --   & 24.0 &--  & 1.29 &1.08 & -0.17  \\

\hline
        \end{array}
    $$
\end{table*}

  \begin{table*}[!ht]
\tiny
     \caption[]{The lines used in the analysis of SDSS J100921 and SDSS J171422 and
     corresponding abundances for models of $T_{\rm eff} = 5820$ K, $\log$ g = 3.9,
     $\xi$ = 1.1 and
    $T_{\rm eff} = 6320$ K, $\log$ g = 4.1, $\xi$ = 1.5, respectively.}
        \label{abund}
    $$
        \begin{array}{l@{}c|@{}r@{}r|@{}c@{}c|@{}cr|| l@{}c|@{}r@{}r|@{}c@{}c|@{}cr}
           \hline
           \hline
          &        & J100921  & J171422 & J100921  & J171422 &      &  &      &        & J100921  & J171422 & J100921 & J171422 &      &     \\
\cline{3-6} \cline{11-14}
 $Species$&\lambda &  $EW$\,\,& $EW$\,\,&\log\epsilon(X)\,\,&\log\epsilon(Fe)\,\,&$LEP$ &\log(gf) &$Species$&\lambda &  $EW$\,\,& $EW$\,\,&\log\epsilon(X)\,\,&\log\epsilon(Fe)\,\,&$LEP$ &\log(gf)\\
\cline{3-6} \cline{11-14}
          & ($\AA$)& (m$\AA$)& (m$\AA$)&($dex$) &($dex$)&($eV$)&   &  &($\AA$)& (m$\AA$)& (m$\AA$)&($dex$)  &($dex$)&($eV$)&    \\
           \hline
Ni$\,{\sc i}$ &4998.23& 14.0 &11.0  &5.16& 5.32 & 3.61& -0.78 & Y$\,{\sc ii}$ &5200.42& SS   &14.0&\leq0.69&1.29&0.99 & -0.57  \\
Ni$\,{\sc i}$ &5000.34& 26.0 &28.0  &5.19& 5.51 & 3.63& -0.43 & Y$\,{\sc ii}$ &5205.73& SS   &SS    &0.79& 1.41 &1.03 & -0.34  \\
Ni$\,{\sc i}$ &5017.54& 41.0 &38.0  &5.12& 5.29 & 3.54& -0.08 & Zr$\,{\sc ii}$&4208.97& 21.0 &20.0  &1.44& 1.73 & 0.71& -0.51  \\
Ni$\,{\sc i}$ &5035.34& 43.0 &53.0  &4.85& 5.29 & 3.63&  0.29 & Ba$\,{\sc ii}$&4554.07& 97.0 &133.0 &0.94& 1.76 &0.00 &  0.17  \\
Ni$\,{\sc i}$ &5042.15& 14.0 &15.0  &5.00& 5.31 & 3.66& -0.57 & Ba$\,{\sc ii}$&5853.68& 29.0 & 42.0 &0.91& 1.41 &0.60 & -1.01  \\
Ni$\,{\sc i}$ &5048.85& 10.0 &18.0  &4.80& 5.37 & 3.85& -0.37 & Ba$\,{\sc ii}$&6141.73& 70.0 & 91.0 &1.07& 1.64 &0.70 & -0.07  \\
Ni$\,{\sc i}$ &5080.53& 44.0 &57.0  &5.05& 5.54 & 3.65&  0.13 & Ba$\,{\sc ii}$&6496.90& 58.0 & 88.0 &0.99& 1.82 &0.60 & -0.41  \\
Ni$\,{\sc i}$ &5081.11& 35.0 &51.0  &4.87& 5.43 & 3.85&  0.30 &               &       &      &      &    &      &     &        \\
\hline				
\hline
        \end{array}
    $$
\end{table*}

  \begin{table*}
\tiny
     \caption[]{The lines used in the analysis of SDSS J015717.04+135535.9 for a models of $T_{\rm eff} = 6250$
     K, $\log$ g = 3.7,
     $\xi$ = 4.0.}
        \label{abund}
    $$
        \begin{array}{l@{}c|rc|cr||l@{}c|rc|cr}
           \hline
\hline
  $Species$&\lambda &  $EW$  &\log\epsilon(X)\,\,& $LEP$ & \log(gf)&$Species$&\lambda &  $EW$  &\log\epsilon(X)\,\,& $LEP$ & \log(gf)\\
\cline{3-6}
\cline{9-12}
           & ($\AA$)&(m$\AA$)& ($dex$)           & ($eV$)&         &         & ($\AA$)&(m$\AA$)& ($dex$)           & ($eV$)&         \\
\hline
C$\,{\sc i}$  & 9094.83 &   SS    &  8.46 & 7.49  & 0.15 &   Fe$\,{\sc i}$ & 4994.13 &   85.0   &  6.55 &  0.92 & -3.07 \\ 
O$\,{\sc i}$  & 7771.94	&   SS    &  8.73 & 9.15  & 0.37 &   Fe$\,{\sc i}$ & 5012.07 &   SS	&  6.32 &0.86	& -2.64 \\ 
O$\,{\sc i}$  & 8446.25	&   SS    &  8.63 & 9.52  &-0.46 &   Fe$\,{\sc i}$ & 5049.81 &   SS	&  6.32 &2.28	& -1.35 \\ 
O$\,{\sc i}$  & 8446.36	&   SS    &  8.63 & 9.52  & 0.24 &   Fe$\,{\sc i}$ & 5051.63 &   SS	&  6.57 &0.92	& -2.78 \\ 
O$\,{\sc i}$  & 8446.76	&   SS    &  8.63 & 9.52  & 0.01 &   Fe$\,{\sc i}$ & 5074.75 &   83.0   &  6.62 &  4.22 & -0.16 \\ 
Mg$\,{\sc i}$ & 5711.10	&   SS    &  6.83 & 4.35  &-1.67 &   Fe$\,{\sc i}$ & 5090.77 &   70.0   &  6.72 &  4.26 & -0.36 \\ 
Al$\,{\sc i}$ & 3944.02 &   SS    &\leq4.27&0.00  &-0.64 &   Fe$\,{\sc i}$ & 5123.72 &   SS	&  6.57 &1.01	& -3.07 \\ 
Al$\,{\sc i}$ & 3961.52 &   SS    &  4.27 & 0.01  &-0.34 &   Fe$\,{\sc i}$ & 5125.11 &   SS	&  6.50 &4.22	& -0.08 \\ 
Si$\,{\sc i}$ & 5708.40	&   SS	  &\leq 6.75&4.95 &-1.37 &   Fe$\,{\sc i}$ & 5127.36 &   SS	&  6.48 &0.92	& -3.31 \\ 
Si$\,{\sc i}$ & 5948.54	&   SS    &  6.80 &5.08   &-1.13 &   Fe$\,{\sc i}$ & 5133.68 &  115.0   &  6.54 &  4.18 &  0.20 \\ 
Ca$\,{\sc i}$ & 4425.44	&   SS    &  5.21 &1.88   &-0.36 &   Fe$\,{\sc i}$ & 5162.27 &  136.0   &  6.86 &  4.18 &  0.08 \\ 
Ca$\,{\sc i}$ & 4434.96 &   SS    &  5.36 &1.89   &-0.01 &   Fe$\,{\sc i}$ & 5202.31 &  100.0   &  6.61 &  2.18 & -1.84 \\ 
Ca$\,{\sc i}$ & 4435.69 &   SS    &  5.36 &1.89   &-0.52 &   Fe$\,{\sc i}$ & 5232.92 &  170.0   &  6.26 &  2.94 & -0.08 \\ 
Ca$\,{\sc i}$ & 4454.78 &   SS    &  5.56 &1.90   & 0.26 &   Fe$\,{\sc i}$ & 5302.31 &   97.0   &  6.48 &  3.28 & -0.74 \\ 
Ca$\,{\sc i}$ & 4455.89 &   SS    &  5.46 &1.90   &-0.53 &   Fe$\,{\sc i}$ & 5332.90 &   80.0   &  6.79 &  1.56 & -2.78 \\ 
Ca$\,{\sc i}$ & 4581.40 &   SS    &  5.56 &2.52   &-0.34 &   Fe$\,{\sc i}$ & 5353.38 &   72.0   &  6.92 &  4.10 & -0.68 \\ 
Ca$\,{\sc i}$ & 4581.47 &   SS    &  5.56 &2.52   &-1.26 &   Fe$\,{\sc i}$ & 5367.44 &   SS	&  6.52 &4.42	&  0.44 \\ 
Ca$\,{\sc i}$ & 4585.87 &   SS    &  5.36 &2.53   &-0.19 &   Fe$\,{\sc i}$ & 5369.93 &   SS	&  6.42 &4.37	&  0.54 \\ 
Ca$\,{\sc i}$ & 4585.96 &   SS    &  5.36 &2.53   &-1.26 &   Fe$\,{\sc i}$ & 5373.71 &   SS	&\leq6.85&4.47  & -0.71 \\ 
Ca$\,{\sc i}$ & 5857.45 &   98.0  &  5.51 &2.93   & 0.23 &   Fe$\,{\sc i}$ & 5383.37 &  114.0   &  6.19 &  4.31 &  0.65 \\ 
Ca$\,{\sc i}$ & 6102.72 &  117.0  &  5.76 &1.88   &-0.79 &   Fe$\,{\sc i}$ & 5393.16 &   SS	&  6.22 &3.24	& -0.72 \\ 
Ca$\,{\sc i}$ & 6122.21 &  134.0  &  5.46 &1.89   &-0.32 &   Fe$\,{\sc i}$ & 5405.77 &  239.0   &  7.02 &  0.99 & -1.85 \\ 
Ca$\,{\sc i}$ & 6162.16 &   SS    &  5.36 &1.90   &-0.09 &   Fe$\,{\sc i}$ & 5429.70 &  189.0   &  6.45 &  0.96 & -1.88 \\ 
Ca$\,{\sc i}$ & 6439.07 &  149.0  &  5.41 &2.53   & 0.47 &   Fe$\,{\sc i}$ & 5432.95 &   SS	&  6.95 &4.45	& -1.02 \\ 
Ca$\,{\sc i}$ & 6449.81 &   73.0  &  5.66 &2.52   &-0.55 &   Fe$\,{\sc i}$ & 5434.52 &  142.0   &  6.21 &  1.01 & -2.12 \\ 
Ca$\,{\sc i}$ & 6462.57 &  140.0  &  5.46 &2.52   & 0.31 &   Fe$\,{\sc i}$ & 5445.04 &  131.0   &  7.03 &  4.39 &  0.04 \\ 
Ca$\,{\sc i}$ & 6471.66 &   SS    &  5.41 &2.53   &-0.59 &   Fe$\,{\sc i}$ & 5446.92 &   SS	&  6.44 &0.99	& -1.91 \\ 
Sc$\,{\sc ii}$& 4246.83 &   SS    &  2.20 &0.32   & 0.24 &   Fe$\,{\sc i}$ & 5473.90 &   SS	&  6.50 &4.15	& -0.79 \\ 
Sc$\,{\sc ii}$&4314.08  &   SS    &  2.15 &0.62   &-0.10 &   Fe$\,{\sc i}$ & 5487.74 &   72.0   &  7.08 &  4.32 & -0.65 \\ 
Sc$\,{\sc ii}$&4320.73  &   SS    &  2.50 &0.61   &-0.25 &   Fe$\,{\sc i}$ & 5497.52 &   SS	&  6.57 &1.01	& -2.84 \\ 
Sc$\,{\sc ii}$&4670.40  &   SS    &\leq2.20&1.36  &-0.58 &   Fe$\,{\sc i}$ & 5501.47 &   SS	&  6.62 &0.96	& -3.05 \\ 
Sc$\,{\sc ii}$&5030.98  &   SS    & 2.15  &1.36   &-0.40 &   Fe$\,{\sc i}$ & 5572.85 &  182.0   &  6.97 &  3.40 & -0.28 \\ 
Sc$\,{\sc ii}$&5526.79  &   SS    &\leq2.30&1.77  & 0.02 &   Fe$\,{\sc i}$ & 5576.08 &   SS	&  6.52 &3.43	& -0.85 \\ 
Ti$\,{\sc i}$ & 4535.92 &   SS    &  3.99 &0.82   &-0.07 &   Fe$\,{\sc i}$ & 5615.62 &  175.0   &  6.51 &  3.33 &  0.05 \\ 
Ti$\,{\sc i}$ & 4991.07 &   SS    &  4.09 &0.84   & 0.38 &   Fe$\,{\sc i}$ & 5762.96 &   SS	&  6.42 &4.21	& -0.41 \\ 
Ti$\,{\sc i}$ & 4999.50 &   SS    &  3.99 &0.83   & 0.25 &   Fe$\,{\sc i}$ & 5930.18 &   SS	&  6.62 &4.65	& -0.23 \\ 
Ti$\,{\sc i}$ & 5129.15 &   SS    &  4.19 &1.89   &-1.40 &   Fe$\,{\sc i}$ & 6024.05 &   90.0   &  6.87 &  4.55 & -0.06 \\ 
Ti$\,{\sc ii}$& 4450.44 &  117.0  &  3.89 &1.08   &-1.45 &   Fe$\,{\sc i}$ & 6065.49 &   SS	&  6.52 &2.61	& -1.53 \\ 
Ti$\,{\sc ii}$& 4468.49 &   SS    &  3.69 &1.13   &-0.62 &   Fe$\,{\sc i}$ & 6230.72 &   88.0   &  6.25 &  2.56 & -1.28 \\ 
Ti$\,{\sc ii}$& 4501.23 &  161.0  &  3.69 &1.12   &-0.75 &   Fe$\,{\sc i}$ & 6252.57 &   78.0   &  6.45 &  2.40 & -1.72 \\ 
Ti$\,{\sc ii}$& 4533.96 &   SS    &  3.99 &1.24   &-0.77 &   Fe$\,{\sc i}$ & 6393.61 &   89.0   &  6.28 &  2.43 & -1.43 \\ 
Ti$\,{\sc ii}$& 4563.72 &  160.0  &  3.99 &1.22   &-0.96 &   Fe$\,{\sc i}$ & 6411.63 &   SS	&  6.88 &3.65	& -0.66 \\ 
Ti$\,{\sc ii}$& 4571.94 &  158.0  &  3.84 &1.57   &-0.52 &   Fe$\,{\sc i}$ & 6597.56 &   SS	&  6.77 &4.80	& -0.92 \\ 
Ti$\,{\sc ii}$& 4589.96 &   SS    &  3.84 &1.24   &-1.78 &   Fe$\,{\sc ii}$& 4508.28 &  120.0   &  6.46 &  2.86 & -2.44 \\ 
Ti$\,{\sc ii}$& 4779.98 &   55.0  &  4.09 &2.05   &-1.37 &   Fe$\,{\sc ii}$& 4520.21 &  121.0   &  6.63 &  2.81 & -2.65 \\ 
Ti$\,{\sc ii}$& 4805.10 &   75.0  &  4.04 &2.06   &-1.12 &   Fe$\,{\sc ii}$& 4522.64 &  169.0   &  6.81 &  2.84 & -2.25 \\ 
Ti$\,{\sc ii}$& 5185.90 &   SS    &  4.09 &1.89   &-1.35 &   Fe$\,{\sc ii}$& 4576.33 &   61.0   &  6.45 &  2.84 & -2.95 \\ 
Ti$\,{\sc ii}$& 5381.01 &   SS    &  4.09 &1.57   &-2.08 &   Fe$\,{\sc ii}$& 4620.52 &   SS	&\leq 6.47&2.83 & -3.21 \\ 
Cr$\,{\sc i}$ & 4646.15 &   SS    &  4.62 &1.03   &-0.71 &   Fe$\,{\sc ii}$& 4629.34 &   SS	&  6.50 &2.81	& -2.38 \\ 
Cr$\,{\sc i}$ & 5204.51 &   SS    &  4.22 &0.94   &-0.21 &   Fe$\,{\sc ii}$& 5234.62 &   SS	&  6.55 &3.22	& -2.18 \\ 
Cr$\,{\sc i}$ & 5208.43 &  169.0  &  4.47 &0.94   & 0.16 &   Co$\,{\sc i}$ & 3995.30 &  112.0   &  3.92 &0.92   & -0.22 \\
Cr$\,{\sc i}$ & 5298.27 &   SS    &  4.72 &0.98   &-1.16 &   Co$\,{\sc i}$ & 4121.28 &   SS	&\leq3.77&0.92  & -0.32 \\
Cr$\,{\sc ii}$& 4558.66 &   86.0  &  4.67 &4.07   &-0.66 &   Ni$\,{\sc i}$ & 5115.37 &   SS	&\leq 5.35&3.83 & -0.11 \\
Cr$\,{\sc ii}$& 4588.18 &   74.0  &  4.52 &4.07   &-0.64 &   Ni$\,{\sc i}$ & 5155.77 &   SS	&\leq 5.38& 3.90& -0.09 \\
Mn$\,{\sc i}$ & 4030.75 &   SS    &  4.49 &0.00   &-0.48 &   Ni$\,{\sc i}$ & 4604.97 &   SS	&  5.50 &3.48	& -0.29 \\
Mn$\,{\sc i}$ & 4033.07 &  209.0  &  4.54 &0.00   &-0.62 &   Ni$\,{\sc i}$ & 4714.42 &   98.0   &  5.30 &3.38   &  0.23 \\
Mn$\,{\sc i}$ & 4034.49 &   SS    &  4.09 &0.00   &-0.81 &   Ni$\,{\sc i}$ & 4715.78 &   SS	&\leq5.25&3.54  & -0.33 \\
Mn$\,{\sc i}$ & 4035.73 &   90.0  &  4.59 &2.14   &-0.19 &   Ni$\,{\sc i}$ & 4786.54 &   74.0   &  5.50 &3.42   & -0.17 \\
Mn$\,{\sc i}$ & 4048.74 &   84.0  &  4.49 &2.16   &-0.13 &   Ni$\,{\sc i}$ & 5146.48 &   64.0   &  5.35 &3.70   &  0.12 \\
Mn$\,{\sc i}$ & 4754.04 &   67.0  &  4.34 &2.27   &-0.09 &   Y$\,{\sc ii}$ & 4883.69 &   SS	&\leq 0.99& 1.08&  0.07 \\
Mn$\,{\sc i}$ & 4823.51 &  113.0  &  4.59 &2.31   & 0.14 &   Ba$\,{\sc ii}$& 5853.68 &   SS	&  1.23 &0.60	& -0.91 \\
Mn$\,{\sc i}$ & 6013.49 &   SS    &\leq4.49&3.07  &-0.25 &   Ba$\,{\sc ii}$& 6141.73 &   SS	&  1.13 &0.70	& -0.03 \\
Mn$\,{\sc i}$ & 6016.64 &   SS    &\leq4.44&3.07  &-0.09 &   Ba$\,{\sc ii}$& 6496.90 &   SS	&  1.18 &0.60	& -0.41 \\
\hline
\hline

       \end{array}
   $$
 \end{table*}

 \begin{table*}[!ht]
\tiny
    \caption[]{Abundances of the observed species for SDSS J100921, J171422, and J015717 
are presented
    for the model atmospheres of $T_{\rm eff} = 5820$ K, $\log$ g = 3.9, $\xi$ = 1.1 and
    $T_{\rm eff} = 6320$ K, $\log$ g = 4.1, $\xi$ = 1.5, and $T_{\rm eff} = 6250$ K, $\log$ g = 3.7, $\xi$ = 4.0, respectively.
    The solar abundances from
    Asplund et al. (2009) are used to
    convert the abundance of element X to [X/Fe].}
       \label{}
   $$
       \begin{array}{l|ccc|ccc|ccc|c}
          \hline
          \hline
 $Species$     &  &J100921 & & &J171422& & &J015717& & \log\epsilon_{\odot}  \\
\cline{2-2}
\cline{3-3}
\cline{4-4}
\cline{5-5}
\cline{6-6}
\cline{7-7}
\cline{8-8}
\cline{9-9}
\cline{10-10}
&\log\epsilon(X) & $[X/H]$ & $[X/Fe]$ & \log\epsilon(X) & $[X/H]$ & $[X/Fe]$ &\log\epsilon(X) & $[X/H]$ & $[X/Fe]$ &  \\
          \hline
 Li$\,{\sc i}$  &1.76 & 0.71&+2.01 &\leq 1.46&\leq0.41 &\leq+1.21&\leq2.21&\leq1.16 &\leq+2.10 &1.05  \\
 C$\,{\sc i}$   &7.58 &-0.85&+0.45  &7.99& -0.44 &+0.36  & 8.46&+0.03  &+0.97&8.43  \\
 O$\,{\sc i}$   &8.28 &-0.41&+0.89  &8.50& -0.19 &+0.61  & 8.68&-0.01  &+0.93&8.69  \\
 Na$\,{\sc i}$  &5.20 &-1.04&+0.26  &5.76& -0.48 &+0.32  & --- &---    &---  &6.24  \\
 Mg$\,{\sc i}$  &6.56 &-1.04&+0.26  &6.88& -0.72 &+0.08  & 6.83&-0.77  &+0.17&7.60  \\
 Al$\,{\sc i}$  &5.07 &-1.38&-0.08  &5.45& -1.00 &-0.20  & 4.27&-2.18  &-1.24&6.45  \\
 Si$\,{\sc i}$  &6.45 &-1.06&+0.24  &6.79& -0.72 &+0.08  & 6.80&-0.71  &+0.23&7.51  \\
 Ca$\,{\sc i}$  &5.18 &-1.16&+0.14  &5.53& -0.81 &-0.01  & 5.46&-0.88  &+0.06&6.34  \\
 Sc$\,{\sc ii}$ &2.06 &-1.09&+0.21  &2.50& -0.65 &+0.15  & 2.25&-0.90  &+0.04&3.15  \\
 Ti$\,{\sc i}$  &3.87 &-1.08&+0.22  &4.36& -0.59 &+0.21  & 4.07&-0.88  &+0.06&4.95  \\
 Ti$\,{\sc ii}$ &4.08 &-0.87&+0.43  &4.33& -0.62 &+0.18  & 3.93&-1.02  &-0.08&4.95  \\
 V$\,{\sc i}$   &2.59 &-1.34&-0.04  &3.47& -0.46 &+0.34  & --- &---    &---  &3.93  \\
 V$\,{\sc ii}$  &2.82 &-1.11&+0.19  &3.42& -0.51 &+0.29  & --- &---    &---  &3.93  \\
 Cr$\,{\sc i}$  &4.32 &-1.32&-0.02  &4.80& -0.84 &-0.04  & 4.51&-1.13  &-0.19&5.64  \\
 Cr$\,{\sc ii}$ &4.48 &-1.16&+0.14  &4.87& -0.77 &+0.03  & 4.60&-1.04  &-0.10&5.64  \\
 Mn$\,{\sc i}$  &3.91 &-1.52&-0.22  &4.56& -0.87 &-0.07  & 4.45&-0.98  &-0.04&5.43  \\
 Fe$\,{\sc i}$  &6.20 &-1.30&+0.00  &6.70& -0.80 &+0.00  & 6.56&-0.94  &+0.00&7.50  \\
 Fe$\,{\sc ii}$ &6.20 &-1.30&+0.00  &6.65& -0.85 &-0.05  & 6.57&-0.93  &+0.01&7.50  \\
 Co$\,{\sc i}$  &3.98 &-1.01&+0.29  &4.34& -0.65 &+0.15  & 3.92&-1.07  &-0.13&4.99  \\
 Ni$\,{\sc i}$  &4.99 &-1.23&+0.07  &5.42& -0.80 &+0.00  & 5.41&-0.81  &+0.13&6.22  \\
 Zn$\,{\sc i}$  &3.36 &-1.20&+0.10  &3.69& -0.87 &-0.07  & --- &---    &---  &4.56  \\
 Sr$\,{\sc ii}$ &1.58 &-1.29&+0.01  &2.11& -0.76 &+0.04  & --- &---    &---  &2.87  \\
 Y$\,{\sc ii}$  &0.87 &-1.34&-0.04  &1.36& -0.85 &-0.05  &\leq 0.99&&\leq-0.28&2.21 \\
 Zr$\,{\sc ii}$ &1.44 &-1.14&+0.16  &1.73& -0.85 &-0.05  & --- &---    &---  &2.58  \\
 Ba$\,{\sc ii}$ &0.98 &-1.20&+0.10  &1.66& -0.52 &+0.28  & 1.18&-1.00  &-0.06&2.18  \\

\hline
\hline
       \end{array}
   $$
\begin{list}{}{}
 \item  Notes. The solar abundances ($\log\epsilon_{\odot}$) are from Asplund et al. (2009).
 \end{list}
  \end{table*}

\section{DISCUSSION}

We performed the first detailed abundance analysis on SDSS J100921 and J171422 for 21 elements and on J015717 for
19 elements. A summary of the abundances for the stars is given in Table 6, where the quantities log
$\epsilon$(X), [X/H], and [X/Fe] are reported in columns two, three, and four. For the SDSS J100921 and J171422, these
abundances were computed for the common lines with the model parameters reported in Table 1. Table 2, 3, and 4
list the elemental abundances for these two program stars while the abundances for the J015717 is listed in Table
5.

In this analysis, the stars, J171422 and J015717, are found to be $\alpha$-poor stars, i.e., the
[$\alpha$/Fe] ratio is 0.09 for J171422 and 0.13 for J015717. Their silicon abundances agree within 0.01
dex. The $\alpha$-poor characteristics of these two {\sc SDSS} stars are not common in the literature.
The stars, BD\,+80\,245, G4-36, and CS\,22966-043 are such low-$\alpha$ halo stars with unusually low
abundances of $\alpha$- (Mg, Si, Ca) and neutron-capture (Sr, Y, Ba) elements (Ivans et al. 2003). The
spectrum of J100921 with $\alpha$-element/Fe]=0.22 is slightly enriched in $\alpha$-elements.    

The mean [$\alpha$/Fe] values calculated over seven (low-$\alpha$) halo stars of Nissen \&
Schuster (1997) are ranging between 0.11 - 0.24. The [$\alpha$/Fe] ratio for the star,
J100921 is, apparently, close to that upper limit within 0.06 dex. The same ratio for J015717
is in the range.
Over twenty-three turnoff halo stars of Nissen et al. (2005), we find a mean value of
[Ca/Fe]=0.25 dex and define a range in [Ca/Fe] ratio: 0.11 $\le$ [Ca/Fe] $<$ 0.38 (see
their Table 1). The stars, J171422 ([Ca/Fe]=$-$0.01) and J015717 ([Ca/Fe]=0.06) have
apparently solar and near solar Ca abundances ratios, while J100921 with [Ca/Fe]=0.14 is found to be
slightly enriched in Ca. Also, in the [Mg/Fe], [Si/Fe], and [Ti/Fe] vs. [Fe/H] planes
(Figures 1 and 2 in the same paper) of Nissen \& Schuster (2008), J171422 and J015717 appear
to be located in a region dominated by low-$\alpha$ stars with a decreasing trend of
[$\alpha$/Fe] as a function of increasing [Fe/H]. The star, J100921, with [Fe/H]=-1.3,
appears to be beyond the region where the halo and thick-disk stars tend to merge at
[Fe/H]$\approx-$1.2. This star with [Mg/Fe]=0.26 also seems to lie just on the dividing line
for high- and low-$\alpha$ populations in Figure 1 of Nissen \& Schuster (2010) and with
[$\alpha$/Fe]=0.22, it is in the same region as low-$\alpha$ stars in the lower panel of the
same figure while the stars, J171422 and J015717 clearly appear to belong to the
low-$\alpha$  population. The $\alpha$-element abundances for apparently normal halo star
J100921 are in good accord with mean abundances ratios to iron of Mg, Si, Ca, and Ti over
seventeen thick-disk stars of Nissen \& Schuster (2010): 0.34$\pm$0.02, 0.31$\pm$0.05,
0.29$\pm$0.04, and 0.25$\pm$0.03, respectively, within the uncertainties. Also, the star is located
in the region populated by thick disk stars in Figure 12 of Reddy et al. (2006). Estimated [Ni/Fe] ratio for J100921 in Figure 2 of
Nissen \& Schuster (2010) with [Na/Fe]=0.26 is in excellent agreement with the star's
reported [Ni/Fe] ratio in Table 6. Accordingly, estimated [Na/Fe] ratio for J171422 with
[Ni/Fe]=0.00 in the Figure 2 of the same paper is $\approx$0.3 dex and in good agreement with the star's reported
[Na/Fe] ratio.   

It is obvious that [Ca/Fe] and [Ti/Fe] ratios for J171422 and J015717 do not agree with the
idea of the stars being members of a thick-disk population but support the low-$\alpha$
membership for the two stars. The estimated [Na/Fe] ratio for J015717 from Figure 1 of
Nissen \& Schuster (2010) with [Ni/Fe]=0.13 is in very good accord with J100921 and J171422.

Errors in model parameters do not change position of three {\sc SDSS} stars dramatically
but, for instance, shifting J171422 and J015717 into the regions populated by low-$\alpha$
stars in Fig.5 and Fig.6 of Nissen \& Schuster (2010) for Si, Ca, Ti, Na, Cr, i.e., an
increase of 125 K in T$_{\rm eff}$ of the J171422 and J015717 cause increase in abundances
of Ca, Ti, Cr, Ni in the range 0.07 -- 0.12 dex.   

The [$\alpha$-element/Fe] ratios over thirty-six low-$\alpha$ stars in the same
study with [Mg/Fe]=0.14$\pm$0.06, [Si/Fe]=0.15$\pm$0.05, [Ca/Fe]=0.23$\pm$0.05, and
[Ti/Fe]=0.14$\pm$0.07 not only support the low-$\alpha$ membership for J171422 and J015717
but also reveal $\approx$0.2 dex lower Ca abundances for the two stars. 

 \begin{figure}
 \centering
 \includegraphics*[width=9cm,height=8cm,angle=0]{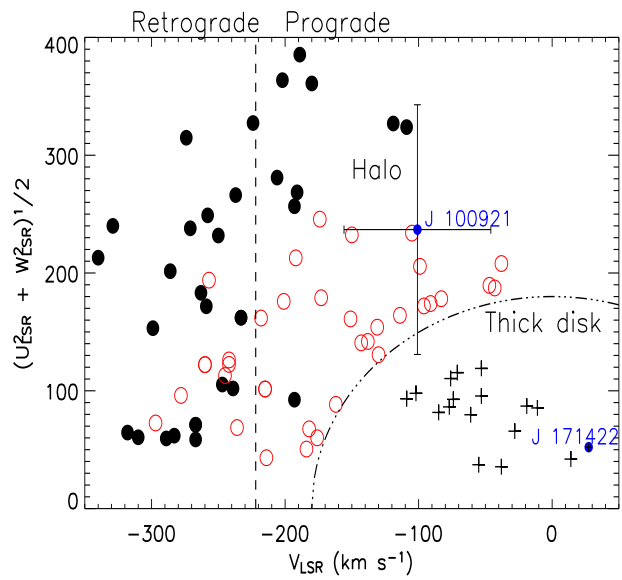}
   \caption{Toomre diagram for stars with [Fe/H] $>$ -1.4. High-$\alpha$ halo stars
are shown with open (red) circles, low-$\alpha$ halo stars with filled (black)
circles, and thick-disk stars with crosses. The dot-dashed line corresponds
to Vtotal = 180 km s$^{-1}$. The vertical dashed line indicates zero rotation
in the Galaxy.}
      \label{toomre}
 \end{figure}

Besides the Nissen et al. studies summarized above, average abundances ratios to iron of Mg, Si, Ca,
and Ti in halo stars in several selected studies, with an average value of [$\alpha$/Fe]=+0.35 over
these four species, ([Mg/Fe]=+0.36, [Si/Fe]=+0.38, [Ca/Fe]=+0.38, and [Ti/Fe]=+0.29; Magain 1987, 1989;
Gratton \& Sneden 1987, 1988, 1991; Zhao \& Magain 1990; Ryan et al. 1991; Nissen et al. 1994; Fuhrmann
et al. 1995; McWilliam et al. 1995) indicates typical $\alpha$-element
abundances for J100921. However, when the thick-disk sample of Reddy et al. (2006) is scrutinized, with
exceptions of too high [O/Fe] and too low [Al/Fe] ratios, the star is clearly seen to resemble a
thick-disk star. The unusually high O abundance of the star is indicated by the O\,{\sc i} lines at
7770 \AA\ and 8440 \AA\ . The former lines will probably be affected by departures from {\sc LTE}
(e.g., Kiselman 1993). Fe contribution to latter group of lines is checked during synthesis analysis
and seen to be negligible.      

We searched for further evidence for halo and/or thick-disk memberships of the program stars using their kinematics. We calculate Galactic velocities ($U$, $V$, $W$; Johnson
\& Soderblom 1987) for the stars, where $U$ is directed towards the Galactic Centre, $V$ is
directed in the sense of Galactic rotation (clockwise as seen from the North Galactic
Pole), and $W$ is directed towards the North Galactic Pole,  with proper motions and
parallaxes taken from the Tycho-2 catalog (Hog et al. 2000, J171422) and the new reduction
of the Hipparcos data (van Leeuwen 2007, J100921). In calculation of the velocities, $U_{LSR}$,
$V_{LSR}$, and $W_{LSR}$, relative to the local standard of rest (LSR), a solar motion of ($U$, $V$, $W$)=(+7.5,+13.5,+6.8) km s$^{-1}$, as derived by
Francis \& Anderson (2009) was assumed. The errors in the parallaxes are $\approx$50$\%$ for J100921
and larger than $\approx$90$\%$ for 171422. The radial velocities of the stars are derived
from the echelle spectra.   

Fig~\ref{toomre} shows the Toomre diagram for the thick-disk and halo stars from Nissen \& Schuster
(2010) with two program SDSS stars reported in the {\sc SIMBAD} shown with filled (blue) circles. In
the same figure, the halo stars are divided into two groups as the high-$\alpha$ or the low-$\alpha$
population. Relatively big errors on the parallaxes for the two program {\sc SDSS} stars indicates
ambiguous memberships for the stars kinematically.  

\section{CONCLUDING REMARKS}

Two of the stars under scrutiny (SDSS J100921.40+375233.9 and J171422.43+283657.2) appear to lie
close to ultra-low metallicity ([Fe/H]=$-$3.5) population track (Fig~\ref{f1_nuv_g}), with the
caveat that the SDSS photometry for them is saturated. For the third object, SDSS
J015717.04+135535.9, one of the available observations place it in the ultra-low metallicity range,
even though four other observations of the object indicated redder NUV$-g$ colors\footnote{Object
appear to lie close to halo metallicity ([Fe/H]=$-$1.5).}, and in this case both the {\sc GALEX} and
the {\sc SDSS} photometry were flagged as problematic. Despite of the high risk, we have followed
spectroscopically these targets to explore the nature of these stars in detail. 

We report that none of these SDSS stars present ultra-low metallicities. Two of the stars, J171422
and J015717 are found to be two low-$\alpha$ halo stars. An interesting anomaly is noted for
J015717.04+135535.9 which exhibits a high rotational velocity (e.g. 40 km s$^{-1}$), unusual for a 
halo turnoff star. The star, J100921 is roughly consistent with thick-disk membership.


\begin{appendix}
\section{ABUNDANCE ANALYSIS - ELEMENTS AND LINES}
\noindent {\bf Li:} The 6707 \AA\ Li\,{\sc i} is detected in J100921. Lithium is assumed to be purely
$^{7}$Li. Then the synthetic spectrum is computed and matched to the spectrum of the star by adjusting
the Li abundance to fit the 6707 \AA\ line. For the spectrum synthesis, a line list for several $\AA$ngstr\"{o}ms
around 6707 \AA\ was compiled using Kurucz's line list\footnote{\it http://kurucz.harvard.edu}. Fe\,{\sc i} line at 6707.44 \AA\ in the blue wing of the Li
line gives no significant contribution. For J100921, we find log$\epsilon$ (Li)=1.76 and [Li/Fe]$\approx$2.0 dex.
The [Li/Fe] ratios of 1.21 and 2.10 for J171422 and J015717, respectively, are listed in Table 6 as upper
limits.\footnote{Uncertainties in derived lithium abundance are estimated from uncertainties in the
atmospheric parameters:$\Delta$T$_{\rm eff}$=+125 K, $\Delta$log\,g=+0.2 cgs units, $\Delta$ $\xi$=+0.5 km
s$^{-1}$ lead to uncertainties of +0.08, $-$0.01, and $-$0.02.}

\vskip 0.2 cm
\noindent {\bf C:} Detection of C\,{\sc i} lines was sought via the 3s$^3$P$^o$-3p$^3$P multiplet with
lines between 9061 \AA\ and 9112 \AA\ . The detected multiplet lines are presented in Table 3 and Table 5
for three SDSS stars. The $gf$-values are taken from Wiese, Fuhr \& Deter (1996). 
\vskip 0.2 cm

\noindent {\bf O:} For J100921.40+375233.9, an abundance $\log\epsilon$(O) $\simeq 8.3$ fits the O\,{\sc i} triplet
at 7774 \AA\ . The triplet is fitted by an abundance of $\log\epsilon$(O) $\simeq 8.5$ for the J171422.
In the spectrum of J015717, only one member of the triplet being 7772 \AA\ O\,{\sc i} was measurable. The 8446
\AA\ feature is also present and provides abundances reported in Table 6. The $gf$-values are taken
from Wiese, Fuhr \& Deter (1996).
\vskip 0.2 cm

\noindent {\bf Na:} Four Na\,{\sc i} lines were suitable for abundance analysis of J171422 (Table 3). 
The Na\,{\sc i} doublet at 5688 \AA\ is fitted by an abundance of $\log\epsilon$(Na)=5.2 for the J100921. The $gf$-values are taken from C. Froese
Fischer\footnote{The MCHF (multi-configuration Hartree-Fock) collection,   
\it http://www.vuse.vanderbilt.edu/~cff/mchf collection/} who used dipole length and
dipole velocity form of the LS line strength in the calculations of oscillator strengths. Achieved
accuracy between the length and velocity results are quite small, of the order of 0.01 - 0.02.   
For J015717, the Na lines used in determination of abundances for J100921 and J171422 were mostly
smeared out due to high value of the stellar rotation. In addition to smearing affect, some
Na lines (e.g. Na\,{\sc i} lines at 5688 \AA\ and 8195 \AA\ ) showed doubling
while some were found to be heavily blended (e.g. Na\,{\sc i} line at 8183 \AA\ ).

\vskip 0.2 cm

\noindent {\bf Mg:} For the two weak Mg\,{\sc i} lines at 4571 \AA\  (RMT1) and 5711 \AA\  (RMT8), the
$gf$-value are taken from G. Tachiev \& C. Froese
Fischer\footnote{\it http://www.vuse.vanderbilt.edu/~cff/mchf collection/} and Chang \& Tang (1990),
respectively. In the latter source, theoretical oscillator strengths are presented. 
\vskip 0.2 cm

\noindent {\bf Si:}  Silicon is represented solely by the Si\,{\sc i} lines at 5708 \AA\  (RMT10) and
5948 \AA\  (RMT16)
. For the Si lines, the $gf$-values are taken from Allende Prieto et al. (2004).
\vskip 0.2 cm

\noindent {\bf Ca:} The $gf$-values for the Ca\,{\sc i} lines listed in Table 3 and Table 5 are taken
from a variety of sources including Olsen, Routly \& King
(1959) and K\"{o}stlin (1964) and for some, the $gf$-values listed in the {\sc NIST} database are used.
New measurement for the common Ca\,{\sc i} line at 6162 \AA\  (RMT3) by Aldenius et al. (2009) is
smaller by only -0.09 dex. 
\vskip 0.2 cm

\noindent {\bf Sc:} The gf-values for the Sc\,{\sc ii} lines in Table 3 and Table 5 are from Lawler \& Dakin (1989) who combined radiative lifetime and
branching ratio measurements.
\vskip 0.2 cm

\noindent {\bf Ti:} For J100921 and J171422, several neutral and single ionized Ti lines are detected.
The $gf$-values for the Ti\,{\sc i} are taken from the {\sc NIST} database. For the
Ti\,{\sc ii} lines, we take the semi-empirical $gf$-values from Pickering et al. (2001, 2002).
\vskip 0.2 cm

\noindent {\bf V:} Vanadium is represented by the one weak line of RMT21 at 4444 \AA\ 
The EWs for the line for J100921 and J171422 are reported in Table 3. The $gf$-values for those lines are taken from Biemont
et al. (1989). For J015717, no neutral vanadium line was detectable. The ionized V lines around 3950 \AA\ region
appear to be smeared out due to high value of the stellar rotation.
the star. 
\vskip 0.2 cm

\noindent {\bf Cr:} The gf-values for Cr\,{\sc i} lines are taken from the {\sc NIST} database in which
the following references dominate as source of transition probabilities for the Cr lines listed in
Table 3 and Table 5: J. M. Bridges (private communication, 1976), Blackwell et al. (1984), Tozzi,
Brunner \& Huber (1985), and Kostyk (1981).   
Sobeck et al.'s (2007) measurements are within $\pm$0.03 dex of {\sc NIST}
values for the common lines in Table 3 and Table 5. The $gf$-values for Cr\,{\sc ii} lines are from {\sc } NIST database.
\vskip 0.2 cm

\noindent {\bf Mn:} The gf-values for Mn\,{\sc i} lines are taken from Blackwell- Whitehead \&
Bergemann (2007) when available or otherwise from the {\sc NIST} database. Hyperfine structure was
considered for all lines with data taken from Kurucz\footnote{\it http://kurucz.harvard.edu}, as
discussed by Prochaska \& McWilliam (2000).
\vskip 0.2 cm

\noindent {\bf Fe:} The original sources for the transition probabilities of the Fe\,{\sc i} lines are listed by
Lambert et al. (1996). The $gf$ values for Fe\,{\sc ii} lines are taken from Melendez et al. (2006).
This revised Fe\,{\sc ii} line list by Melendez et al. (2006) takes into account new laboratory
measurements by Schnabel et al. (2004) and new theoretical calculations by R. L. Kurucz.\footnote{\it
See http://kurucz.harvard.edu} The $gf$ values reported by Melendez et al. (2006) were put onto the laboratory scale via experimental branching ratios and
radiative lifetimes. The mean difference between the $gf$ values for Fe\,{\sc ii} lines in Lambert et
al. (1996) and Melendez et al. (2006) is 0.03 dex. 
\vskip 0.2 cm

\noindent {\bf Co:}  The search for Co\,{\sc i} lines drew on the tabulation of experimental gf-values provided by
Cardon et al. (1982). The Co\,{\sc i} lines at 3995.31 \AA\ , 4121.32 \AA\ are present.
\vskip 0.2 cm

\noindent {\bf Ni:} A dominant source of the $gf$-values for the Ni\,{\sc i} lines is Kostyk (1982b).
The $gf$-values from Doerr \& Kock (1985) and Lennard et al. (1975) are also used.
\vskip 0.2 cm

\noindent {\bf Zn:} Zinc is represented by the two Zn\,{\sc i} lines of RMT2 at 4722 \AA\ and 4810 \AA\ . The former
line is not detected in J015717 and also, the letter one could not be resolved from the background. The
gf-values are from Bi\'{e}mont \& Godefroid (1980).
\vskip 0.2 cm

\noindent {\bf Sr:} The strong resonance lines of Sr\,{\sc ii} at 4077 \AA\ and 4215 \AA\ are present in
the spectra of J100921 and J171422. Their equivalent widths are listed in Table 4. The $gf$-values are from Brage et al.
(1998).
\vskip 0.2 cm

\noindent {\bf Y:} Selection of Y\,{\sc ii} lines is based on the solar lines judged to be unblended
Y\,{\sc ii} lines in the solar spectrum by Hannaford et al. (1982) who provide accurate gf-values. The
six Y lines used in the analysis of J100921 and J171422 are listed in Table 4. We present the mean Y
abundances for both stars in Table 6.
\vskip 0.2 cm

\noindent {\bf Zr:} Our search for Zr\,{\sc ii} lines drew on the papers by Bi\'{e}mont et al. (1981) and
Ljung et al. (2006) who measured accurate laboratory gf-values and conducted an analysis of Zr\,{\sc
ii} lines to determine the solar Zr abundance. The Zr\,{\sc ii} line at 4208 \AA\ is detected in two of
the program stars: J100921 and J171422. The adopted $gf$-value for the line differs only -0.05 dex
from the $gf$-value of Hannaford et al. (1981). For J015717, no zirconium line was detected.
\vskip 0.2 cm

\noindent {\bf Ba:} The Ba abundance is based on the Ba\,{\sc ii} line at 4554 \AA\ and the lines at
5853, 6141, and 6496 \AA\ . The gf-value adopted is the mean of the experimental values from Gallagher
(1967) and Davidson et al. (1992). Hyperfine and isotopic splittings are taken into account from
McWilliam (1998).

\end{appendix}

\end{document}